\documentclass[twocolumn]{aastex62}

\usepackage{amsfonts,amssymb,amsmath}
\usepackage{comment}
\usepackage{mycommands}
\graphicspath{ {figs/} }

\usepackage{hyperref}
\hypersetup{
	breaklinks=true,
	colorlinks=true,
	linkcolor=blue,
	citecolor=purple,
	urlcolor=purple,
	pdfstartview={FitH},
	pdfview={FitH 0},
	pdfauthor={Navid C. Constantinou and Jeffrey B. Parker},
	pdftitle={Magnetic suppression of zonal flows on a beta plane},
 }

\renewcommand{\eref}[1]{Eq.~\eqref{#1}}
\newcommand{\erefs}[1]{Eqs.~\eqref{#1}}

\newcommand{\J}{\mathsf{J}}
\newcommand{\Alfven}{Alfv{\'e}n}
\newcommand{\wA}{\w_A}  
\newcommand{\wR}{\w_R}  
\newcommand{\wf}{\w_f} 		
\newcommand{\ws}{\w_s}		
\newcommand{\ubar}{\bar{u}}
\newcommand{\Prm}{\mathrm{Pr}_{\mathrm{m}}}
\newcommand{\Y}{\Upsilon}

\received{May 24, 2018}
\revised{June 19, 2018}
\accepted{June 19, 2018}

\submitjournal{ApJ}

\shorttitle{Magnetic suppression of zonal flows on a beta plane}
\shortauthors{Constantinou and Parker}

\begin{document}

\title{Magnetic suppression of zonal flows on a beta plane}

\correspondingauthor{Navid C. Constantinou}
\email{navid.constantinou@anu.edu.au}

\author[0000-0002-8149-4094]{Navid C. Constantinou}
\affiliation{Scripps Institution of Oceanography, University of California San Diego, La Jolla, CA~92093-0213, USA}
\affiliation{Research School of Earth Sciences, Australian National University, Canberra, ACT, 2601, Australia}
\affiliation{ARC Centre of Excellence for Climate Extremes, Australian National University, Canberra, ACT, 2601, Australia}

\author[0000-0002-9079-9930]{Jeffrey B. Parker}
\affil{Lawrence Livermore National Laboratory, Livermore, CA~94550, USA}

\begin{abstract}
Zonal flows in rotating systems have been previously shown to be suppressed by the imposition of a background magnetic field aligned with the direction of rotation.  Understanding the physics behind the suppression may be important in systems found in astrophysical fluid dynamics, such as stellar interiors.  However, the mechanism of suppression has not yet been explained.  In the idealized setting of a magnetized beta plane, we provide a theoretical explanation that shows how magnetic fluctuations directly counteract the growth of weak zonal flows.  Two distinct calculations yield consistent conclusions.  The first, which is simpler and more physically transparent, extends the Kelvin--Orr shearing wave to include magnetic fields and shows that weak, long-wavelength shear flow organizes magnetic fluctuations to absorb energy from the mean flow.  The second calculation, based on the quasilinear, statistical CE2 framework, is valid for arbitrary wavelength zonal flow and predicts a self-consistent growth rate of the zonal flow.  We find that a background magnetic field suppresses zonal flow if the bare {\Alfven} frequency is comparable to or larger than the bare Rossby frequency.  However, suppression can occur for even smaller magnetic fields if the resistivity is sufficiently small enough to allow sizable magnetic fluctuations.  Our calculations reproduce the $\eta/B_0^2 = \text{const.}$ scaling that describes the boundary of zonation, as found in previous work, and we explicitly link this scaling to the amplitude of magnetic fluctuations.
\end{abstract}


\section{Introduction} \label{sec:intro}

Zonal flows, or latitudinal bands of east--west alternating fluid flow, commonly form in the atmospheres of rotating planets \citep{ingersoll:1990,vasavada:2005}.  In contrast, in the solar tachocline, in which a background toroidal magnetic field is present, zonal flows are not commonly thought to occur.  The solar tachocline, the thin layer between the radiative zone and convective zone, may play an important role in the solar dynamo \citep{spiegel:1992,tobias:2002,wright:2016}.  Understanding plasma dynamics under the combined influence of both rotation and magnetic field can help provide insight into the solar tachocline, to other stellar interiors, gas giant interiors, and possibly to exoplanets.

\citet{tobias:2007} studied a two-dimensional (2D) magnetized beta plane as a way to gain insight into how a magnetic field affects turbulence and zonation in a rotating, stratified system.  The magnetized beta plane, while a relatively simple model, contains some of the key physics of the tachocline.  Through direct numerical simulations, they found that when the mean toroidal magnetic field is strong enough, formation of zonal flow is suppressed.

In a follow-up work, \citet{tobias:2011} generalized the numerical simulations from the beta plane to full spherical geometry.  On the surface of a rotating sphere, turning on an azimuthal background magnetic field also suppressed formation of zonal flow.  In that work, the authors did not identify any fundamentally new mechanism of suppression present on a spherical surface that was absent on a beta plane.  In addition to direct numerical simulations, \citet{tobias:2011} showed that the statistical model CE2 captures the zonal-flow-suppression mechanism. CE2 is based on a quasilinear approximation, where the eddy--eddy nonlinearity is neglected from the eddy dynamics but kept intact in the mean flow dynamics (for details regarding CE2 see Section~\ref{sec:CE2formulation}). 

An open question raised by the numerical results of \citet{tobias:2007} and \citet{tobias:2011} is what exactly is the mechanism that suppresses zonal flows.  Their calculations just described have employed nonlinear, time-evolving simulations in which a variety of processes can occur and coexist.  Understanding the detailed physics and mechanisms underlying the suppression of mean zonal flows would be valuable.

Here, we reconsider the 2D magnetized beta plane studied by \citet{tobias:2007} in order to investigate in more detail the suppression of zonal flow.  In the simple geometry of the beta plane, analytic calculations are more tractable than on the sphere.  We adopt a quasilinear approach and use the CE2 statistical framework. The CE2 framework has proven successful in understanding zonal flows on the unmagnetized beta plane \citep{farrell:2007,srinivasan:2012,tobias:2013,constantinou:2014,constantinou:2016,parker:2013,parker:2014}.  CE2 has also been applied in astrophysical fluid dynamics in an MHD setting to study the magnetorotational dynamo \citep{squire:2015}.  Encouragingly, that study found that the quasilinear model qualitatively reproduced the dependence of a key figure of merit on the magnetic Prandtl number $\Prm$.  

Within the CE2 framework, we calculate the eigenvalues and eigenmodes of the linear instability in which zonal flows grow, known as `zonostrophic instability' (see Section~\ref{sec:ZI}).  Zonostrophic instability refers to the process in which a weak zonal flow in an otherwise homogeneous turbulent field organizes the incoherent fluctuations to \emph{coherently} reinforce the zonal flow.  We find that the presence of a background magnetic field suppresses the zonostrophic instability.

Additionally, in Section~\ref{sec:Orr}, we perform a related, but simpler and more physically transparent, calculation based on the Kelvin--Orr shearing wave \citep{thomson:1887,orr:1907}.  Starting with the work by \citet{kraichnan:1976} and then followed with those by \citet{huang:1998}, \citet{chen:2006}, \citet{holloway:2010}, and \citet{cummins:2010}, it has been shown that when \emph{strong} mean flows are present, the Kelvin--Orr shearing wave dynamics is the dominant process by which energy is transferred from the small-scale fluctuations to large-scale mean flows.  However, more recently \citet{bakas:2013-jas} further demonstrated that the Kelvin--Orr shearing wave dynamics can also be important when mean flows are \emph{weak}, since the shearing wave dynamics underlie the organization of incoherent fluctuation to drive mean flows. Here, we extend the weak-mean-flow Kelvin--Orr shearing wave dynamics to include magnetic field. The shearing wave solution we derive demonstrates that while hydrodynamic fluctuations may transfer energy to the mean flow, the magnetic field essentially always counteracts energy transfer to the mean flow. Further, we show that the parameter dependence found in the Kelvin--Orr calculation is recovered by the zonostrophic-instability computation in the appropriate asymptotic regime.

\section{Formulation} \label{sec:formulation}


We consider the quasi-geostrophic dynamics of an incompressible, magnetized fluid on a beta plane $\bx\equiv(x,y)$, with $x$ being the azimuthal direction (longitude) and $y$ the meridional direction (latitude).  A beta plane is a geometrical simplification of a rotating sphere that retains the physics associated with rotation and the latitudinal variation of rotation velocity \citep{pedlosky:book}.  The beta plane uses a Cartesian geometry, and the gradient of the Coriolis parameter is described by a constant parameter $\beta$.  We use periodic boundary conditions in both directions.

The fluid velocity $\v{u} = (u,v)$ derives from a stream function $\psi(\v{x},t)$, i.e., $\v{u} = \unit{z} \v{\times} \v{\nabla} \psi$. The vorticity normal to the plane of motion is $\z \equiv \unit{z} \bcdot (\v{\nabla} \v{\times} \v{u}) = \nabla^2 \psi$.  The magnetic field is given in terms of a vector potential, $\v{B} \equiv \v{\nabla}\v{\times}\v{A}$ and it consists of a constant, uniform background $B_0 \unit{x}$ in the azimuthal direction and a time-varying component, such that $\v{B} \equiv (B_0 + \partial_y A) \unit{x} - (\partial_x A) \unit{y}$, where $\v{A} = [B_0 y + A(\v{x}, t)] \unit{z}$ is the vector potential.

The magnetohydrodynamics (MHD) evolution of the system can be described by a formulation involving vorticity and magnetic potential,
\begin{subequations}
	\label{eq:EOM}
	\begin{gather}
		\partial_t \z + \J(\psi, \z+\beta y) = \J(A + B_0y, \nabla^2 A) + \nu \nabla^2 \z + \xi, \label{eq:psi}\\
		\partial_t A  + \J(\psi, A + B_0y) = \eta \nabla^2 A. \label{eq:A}
	\end{gather}
\end{subequations}
In~\eref{eq:EOM}, $\J(a, b) \equiv (\partial_x a)(\partial_y b) - (\partial_y a)(\partial_x b)$ is the Poisson bracket, $\b$~is the latitudinal gradient of the Coriolis parameter, $\nu$ is the viscosity, $\eta$ is the resistivity, and $\xi(\v{x},t)$ is a random forcing to excite fluctuations.  For mathematical convenience, we have set the permeability $\mu_0 = 1$ and the mass density $\rho=1$.  In these units, the background magnetic field $B_0$ is equivalent to the {\Alfven} velocity $v_A = B_0 / \sqrt{\mu_0 \rho}$.

The first term on the right-hand side of \eref{eq:psi} is the curl of the Lorentz force, $\v{j} \v{\times} \v{B}$.  Equation~\eqref{eq:A} is an expression of Faraday's law combined with Ohm's law, $\v{E} = -\v{u} \v{\times} \v{B} + \eta \v{j}$, and Amp\`{e}re's law, $\v{j} = \v{\nabla} \v{\times} \v{B}$.

In~\eref{eq:psi}, $\xi$~is a stochastic excitation that is assumed \emph{(i)}~to have zero mean (over space, time, or ensemble), \emph{(ii)}~to be spatially and temporally statistically homogeneous, and \emph{(iii)}~to be temporally delta correlated but spatially correlated. Thus, it satisfies,
\begin{subequations}
	\begin{align}
		\langle \xi(\v{x},t) \rangle &= 0, \label{eq:forc1}\\
		\langle \xi(\v{x}_a,t_a)\xi(\v{x}_b,t_b)\rangle &= Q(\v{x}_a-\v{x}_b)\,\delta(t_a-t_b)  ,\label{eq:forc2}
	\end{align}
\end{subequations}
where angle brackets denote ensemble average over different forcing realizations.  The spatially homogeneous forcing can be prescribed by the Fourier spectrum of its covariance through
	\begin{align}
		Q(\v{x}_a - \v{x}_b) = \sum_{\v{k}} \hat{Q}_{\bk}\,e^{i\v{k}\bcdot(\v{x}_a-\v{x}_b)}.
	\end{align}

We observe that in the magnetized beta plane, where the background magnetic field is aligned along the direction of rotation, the resulting dynamics are not dependent on the sign of $B_0$.  To see this, note that we are free to let $A \to -A$ in the definition of $\v{A}$, as this is merely a choice of sign convention.  If we set both $A \to -A$ and $B_0 \to -B_0$ in \eref{eq:EOM}, then the dynamics is unchanged.

\subsection{Fast and slow magneto-Rossby waves}
The system of \eref{eq:EOM} supports two basic waves, the fast and slow magneto-Rossby waves, which are mixtures of the Rossby wave and the shear {\Alfven} wave.  To derive the dispersion relations of the magneto-Rossby waves, we linearize the unforced equations of motion about $(\z, A) = (0, 0)$ and substitute perturbations of the form $e^{i \v{k} \cdot \v{x} - i \w t}$.  We obtain the dispersion relation
	\begin{align}
		\w_{f,s} = \frac{\w_R}{2} \left( 1 - i \frac{(\n + \eta)k^2}{\w_R} \pm \sqrt{\left[ 1 - i\frac{(\n - \eta)k^2}{\w_R} \right]^2 + \frac{4 \w_A^2}{\w_R^2} } \right) ,
		\label{fast_slow_frequencies}
	\end{align}
where $k^2 = k_x^2 + k_y^2$ and $\w_R \equiv -\b k_x/k^2$ and $\w_A \equiv k_x B_0$ are the frequencies of the undamped Rossby and shear {\Alfven} waves, respectively. The fast wave $\w_f$ takes the $+$ sign, and the slow wave $\w_s$ takes the $-$ sign.

The eigenmodes can be obtained from the linearized magnetic equation as
	\begin{equation}
		\begin{bmatrix} \z \\ A \end{bmatrix}_{f,s} = \frac{1}{k \sqrt{|\w_{f,s} + i \eta k^2|^2 + \w_A^2}}  \begin{bmatrix} k^2 (\w_{f,s} + i \eta k^2) \\ -\w_A \end{bmatrix}.\label{eq:eigenmodes_fs}
	\end{equation}
For later convenience, the normalizing factor has been chosen such that the quantity $k^2 (|\psi|^2 + |A|^2) = k^2 ( |\z|^2 / k^4 + |A|^2)$, which is equal to the mode energy (up to a factor of 2), is unity.

We examine two limits to elucidate the physical nature of these waves.  First, in the nondissipative limit where $\n$ and $\eta$ vanish, the frequencies are
	\begin{equation}
		\w_{f,s} = \frac{\wR}{2} \left(1 \pm \sqrt{1 + \frac{4 \wA^2}{\wR^2}} \right).
	\end{equation}
For vanishing magnetic field the Rossby wave is recovered, while for strong magnetic field the shear {\Alfven} wave is recovered.

Second, in this paper we focus on the regime where in the length scales of interest, the Rossby wave is the fastest process, such that $\n k^2$, $\eta k^2$, $\w_A \ll \w_R$.  In this regime, \eref{fast_slow_frequencies} reduces to
\begin{subequations}
	\label{eqs:fastslow_easyregime}
	\begin{align}
		\w_f &= \w_R - i \n k^2,\\
		\w_s &= -\frac{\w_A^2}{\w_R} - i \eta k^2. 
  \end{align}
\end{subequations}
The fast wave is essentially the Rossby wave, while the slow wave involves both the magnetic field and beta effect.  In this regime, the eigenmodes in~\eref{eq:eigenmodes_fs} simplify~to
	\begin{subequations}
		\label{eigenmode_relations_easyregime}
	\begin{align}
		\begin{bmatrix} \z \\ A \end{bmatrix}_f &= \frac{1}{k} \begin{bmatrix} k^2 \\ -\w_A / \w_R \end{bmatrix} , \\
		\begin{bmatrix} \z \\ A \end{bmatrix}_s &= \frac{1}{k} \begin{bmatrix} k^2 \w_A / \w_R \\ 1 \end{bmatrix} .
	\end{align}
	\end{subequations}
In this regime, the fast wave is dominated by the vorticity component and the slow wave is dominated by the magnetic component.

\subsection{Quasilinear dynamics and the CE2 second-order closure}
\label{sec:CE2formulation}
A useful framework for addressing the dynamics of coherent flows embedded in and driven by turbulence involves studying the dynamics of the \emph{statistics} of the flow fields (e.g., statistical moments).  Rather than working directly with flow fields that rapidly vary in time and space, studying the behavior of dynamical equations for statistical quantities can provide qualitative insight of turbulence--mean flow interaction.  However, forming statistically averaged equations of nonlinear systems inevitably runs into the closure problem, where an infinite hierarchy of moment equations is required to obtain a closed system.  Thus, a turbulence closure is needed. 

Here, we study the dynamics of the magnetized fluid in \eref{eq:EOM} using the quasilinear second-order closure.  This closure has proven useful in gaining analytic understanding and physical insight regarding coherent-structure formation in turbulent flows.  In the quasilinear second-order closure, the eddy--mean flow interaction is accurately captured; indeed, this interaction is not approximated whatsoever.  This particular closure comes (unfortunately) in the literature under two names: ``S3T', which stands for Stochastic Structural Stability Theory \citep{farrell:2003} and ``CE2'', which stands for Cumulant Expansion at second order \citep{marston:2008}. Hereafter we refer to this closure as CE2. 

We consider a decomposition of the flow fields into a coherent and an incoherent component. Here, we identify the coherent component with the zonal mean (denoted by over bar) and the incoherent component, or eddies, with the fluctuations about the zonal mean (denoted by prime),~e.g.,
\begin{subequations}
	\begin{align}
		\ol{\psi}(y, t) &\equiv \frac{1}{L_x} \int_0^{L_x} dx\, \psi(\v{x}, t), \\
		\psi'(\v{x}, t) &\equiv \psi(\v{x},t) - \ol{\psi}(y,t).
	\end{align}
\end{subequations}
The quasilinear approximation consists of neglecting the eddy--eddy nonlinearity in the eddy evolution equations while keeping the mean flow dynamics intact. Thus, from \eref{eq:EOM}, we obtain the quasilinear equations
	\begin{subequations}
	\label{eqs:QL2}
	\begin{align}
		\partial_t \bar{u}  &=  \overline{v'\,\z'} - \overline{(\partial_x A')\nabla^2 A'}  + \nu \partial_y^2\bar{u},\label{eq:ubar}\\
		\partial_t \bar{A}  &= - \partial_y(\overline{ v'A'}) + \eta\partial_y^2\bar{A},\label{eq:Abar}\\
		\partial_t \z'  &+ \bar{u}\partial_x\z' + (\beta -\partial^2_y\bar{u}) v'  =\notag \\
			& = -(B_0 + \partial_y\bar{A})\partial_x\nabla^2 A' + (\partial_y^3\bar A)(\partial_x A') + \nu\nabla^2\z' + \xi ,\label{eq:psipql}\\
		\partial_t A'  &+ \bar{u} \partial_x A' = -(B_0 + \partial_y\bar{A})v' + \eta\nabla^2 A' .\label{eq:Apql}
	\end{align}
	\end{subequations}

From the quasilinear equations above we can form the closed system for the evolution of the first and second flow cumulants. The first cumulants being the mean flow components,
	\begin{equation}
		\bar{u}\ \text{and}\ \bar{A},
	\end{equation}
while the second cumulants are the same-time two-point eddy covariances:
	\begin{align}
		W \equiv \ol{ \z'(\v{x}_a,t)\z'(\v{x}_b,t) }, & \quad M \equiv \ol{ \z'(\v{x}_a,t)A'(\v{x}_b,t) }, \notag \\
		N \equiv \ol{ A'(\v{x}_a,t)\z'(\v{x}_b,t) }, & \quad G \equiv \ol{ A'(\v{x}_a,t)A'(\v{x}_b,t) }.\label{eq:def_cov}
	\end{align}
The stresses that appear in the mean flow equations~\eqref{eq:ubar} and~\eqref{eq:Abar} are expressed in terms of the eddy covariances through
	\begin{subequations}
	\label{eq:stresses}
	\begin{align}
		\ol{v'\z'} &= \tfrac1{2}\left[ (\partial_{x_a}\nabla^{-2}_a+\partial_{x_b}\nabla^{-2}_b) W \right]_{a=b},\\
		\ol{(\partial_x A')\nabla^{2} A'} &= \tfrac1{2}\left[ (\partial_{x_a}\nabla^{2}_b + \partial_{x_b}\nabla^{2}_a) G \vphantom{\nabla^{-2}_b}\right]_{a=b}, \\
		\ol{v'A'} &= \tfrac1{2}\left[ \partial_{x_a}\nabla^{-2}_a M + \partial_{x_b}\nabla^{-2}_b N \right]_{a=b},
	\end{align}
	\end{subequations}
where the subscript $a=b$ denotes that the function of $\v{x}_a$ and~$\v{x}_b$ inside square brackets is transformed into a function of a single spatial coordinate by setting~$\v{x}_a=\v{x}_b=\v{x}$. Thus, the mean flow equations in the CE2 closure are exactly~\eref{eq:ubar} and~\eref{eq:Abar} with the stresses given by~\eref{eq:stresses}.

By manipulating~\eref{eq:psipql} and~\eref{eq:Apql} and also using \eref{eq:forc2} we obtain the evolution equations for the eddy covariances~\eref{eq:def_cov}:
	\begin{subequations}
	\label{eqs:covar}
	\begin{align}
		\partial_t W & = \,(\Lc^{\z\z}_a + \Lc^{\z\z}_b ) W \,+\, \Lc^{\z A}_a N  \,+ \Lc^{\z A}_b M + Q  ,\label{eq:Comom}\\
		\partial_t M & = (\Lc^{\z\z}_a + \Lc^{AA}_b) M + \, \Lc^{\z A}_a G \,+ \Lc^{A\z}_b W  ,\label{eq:ComA}\\
		\partial_t G & = (\Lc^{AA}_a+\Lc^{AA}_b) G + \Lc^{A\z}_a M + \Lc^{A\z}_b N\label{eq:CAA} ,
	\end{align}
	\end{subequations}
where the $\Lc$ operators depend on the mean flow fields, $\bar{u}$ and~$\bar{A}$, and are given by
	\begin{subequations}
	\label{eqs:Lops}
	\begin{align}
		\Lc^{\z\z} &\equiv -\bar{u}\,\partial_x - (\beta-\partial^2_y\bar{u})\nabla^{-2}\partial_x+ \nu 	\nabla^{2},\\
		\Lc^{\z A} &\equiv  -(B_0 + \partial_y\bar{A}) \nabla^{2}\partial_x + (\partial^3_y\bar{A})\partial_x,\\
		\Lc^{A\z} &\equiv -(B_0 + \partial_y\bar{A}) \nabla^{-2}\partial_x,\\
		\Lc^{AA} &\equiv -\bar{u}\,\partial_x + \eta\nabla^{2}.
	\end{align}
	\end{subequations}
In~\eref{eqs:covar}, $Q$ is the forcing covariance defined in \eref{eq:forc2} and subscripts in the $\Lc$ operators denote the variables on which the differential operators act and at which the mean flow fields are evaluated.  We have assumed ergodicity to replace zonal averages over the random-forcing realizations with their ensemble averages.  The evolution equation for mixed covariance $N$ is redundant because of the symmetry
	\begin{equation}
		M(\v{x}_a,\v{x}_b,t) = N(\v{x}_b,\v{x}_a,t).\label{eq:symm_MN}
	\end{equation}
The evolution equation for $N$ can be obtained from \eref{eq:ComA} by exchanging $\z\leftrightarrow A$ in the superscripts of the $\Lc$ operators together with exchanging $a\leftrightarrow b$ in the subscripts.

Note that only the quasilinear approximation in \eref{eqs:QL2} is enough produce the CE2 closure. Thus, a closure of the flow statistics at second order is exactly equivalent with the neglect of the eddy--eddy nonlinearity in the eddy dynamics.

The terms on the right-hand-side of \eref{eq:ubar} can be rewritten using integration by parts as
	\begin{subequations}
	\label{eqs:taylor}
	\begin{align}
		\ol{v'\,\z'} &=  - \partial_y \ol{u'v'}, \label{eq:taylor1}\\
		\ol{(\partial_x A')\nabla^2 A'} &= - \partial_y \ol{B'_x B'_y}. \label{eq:taylor2}
	\end{align}
	\end{subequations}
These identities allow the forces in \eref{eq:ubar} to be written in the form of divergence-of-a-stress.  Equation~\eqref{eq:taylor1} is Taylor's identity that relates the vorticity flux with the Reynolds-stress divergence; \eref{eq:taylor2} is analogous to \eref{eq:taylor1} in providing an identity for the vorticity flux associated with the Maxwell stress.  We will use either of the expressions in~\eref{eq:taylor1} interchangeably and refer to them simply as the ``Reynolds stress''; similarly we refer to either of the expressions in \eref{eq:taylor2} as the ``Maxwell stress.''

In summary, the CE2 equations consist of the evolution \eref{eqs:covar} for the eddy covariances, and the evolution \erefs{eq:ubar}--\eqref{eq:Abar} for the zonally averaged flow and magnetic potential (in which the stresses are given by~\eref{eq:stresses}).

\section{Shearing wave dynamics and energy transfers to a weak, long-wavelength shear flow} \label{sec:Orr}
In this section, we show that a relatively simple mechanism underlies the magnetic suppression of zonal flows.  We revisit the Kelvin--Orr shearing wave, which examines the response of a wave to a fixed, long-wavelength shear flow.  We find that in much the same way that a weak shear flow can organize hydrodynamic fluctuations to reinforce itself, a shear flow can also organize magnetic fluctuations to oppose it.  

The Kelvin--Orr shearing wave was originally used to explain the non-modal growth of perturbations on a shear flow \citep{thomson:1887,orr:1907,tung:1983,boyd:1983,farrell:1987}.  In those studies, the shear flow considered had a \emph{finite amplitude}.  This non-modal growth is sometimes referred to as the Kelvin--Orr mechanism.  In the same limit of strong shear flows, \cite{leprovost:2007} investigated the effect that magnetic fields have on turbulent transport in a setup similar to the one we study here.  With a different physical phenomenon in mind, \citet{bakas:2013-jas} combined the hydrodynamic Kelvin--Orr shearing wave with \emph{weak} shear flow to show that a weak shear flow can drain energy from certain waves leading to mean flow growth.

Here, we extend the analysis of the Kelvin--Orr shearing wave in a weak shear flow to include MHD fluctuations.  We show that the magnetic field inhibits energy transfer from eddies to the zonal flows in two ways: \emph{(i)}~it reduces the range of waves that are able to produce reinforcing Reynolds stresses and \emph{(ii)}~it produces Maxwell stresses that oppose zonal flow growth.  First, we review the basic calculation in a system with no beta effect and no magnetic fields.

\subsection{No beta effect and no magnetic field \label{subsec:Orr}}
Here, we demonstrate how the Kelvin--Orr shearing wave gives rise to the tendency for hydrodynamic fluctuations in the presence of a long-wavelength shear flow to transfer energy to the shear flow.  The calculation was presented by \citet{bakas:2013-jas}, which we review here because we use the same techniques when we include a magnetic field in Section~\ref{subsec:Orr_betaB}.

First, we consider the energetics of the mean flow.  The zonally-averaged momentum equation, ignoring magnetic fields and dissipation, is given by~\eref{eq:ubar}: $\partial_t \bar{u} = -\partial_y \ol{u' v'}$. Multiplying by $\bar{u}$ and averaging over~$y$, we obtain
	\begin{equation}
		\d{E_{{\rm ZF}}}{t} = \frac{1}{L_y} \int_0^{L_y} dy\, \ol{u' v'} \partial_y \ol{u},
		\label{orr:zf_energetics_1}
	\end{equation}
where $E_{{\rm ZF}} \equiv \tfrac{1}{L_y} \int_0^{L_y} dy\, \tfrac{1}{2} \ol{u}^2$ is the spatially averaged energy density of the zonal flow and we have neglected boundary terms.

For the rest of this section, we consider the evolution of perturbation vorticity under the assumption of a fixed, linear shear flow $\bar{u} = Sy$.  Unlike a periodic $\ol{u}$, a linear flow appears incompatible with the neglect of boundary terms, a point which we will return to at the end of this section.  The linearized evolution equation for vorticity~is
	\begin{equation}
		(\partial_t + S y \partial_x) \z' = \n \nabla^{2} \z' .
	\end{equation}

As we are interested in studying the emergence of zonal flows, we assume that $\bar{u}$ is very weak. This assumption implies that the shear $S$ is very small, in a manner to be quantified later.  We substitute an ansatz $\z'(\v{x}, t) = Z(t) e^{i \v{k}(t) \bcdot \v{x}}$.  Requiring the coefficients of the terms linear in $x$ and $y$ to vanish, we see that $d k_x/dt = 0$ and $d k_y/dt = -S k_x$.  Hence,
	\begin{subequations}
		\label{eq:shearwavenumbers}
	\begin{align}
		k_x &= \text{constant}, \\
		k_y(t) &= k_{y0} - S k_x t.
	\end{align}
	\end{subequations}
The resulting equation for the amplitude $Z$ can be solved, yielding
	\begin{equation}
		\z' = Z_0 e^{i[k_x x + k_y(t) y]} e^{-\n \int_0^t d\t\, k(\t)^2},
		\label{orr:basic:shearingwavesolution}
	\end{equation}
where $k(t)^2 \defineas k_x^2 + k_y(t)^2$.  Equation~\eqref{orr:basic:shearingwavesolution} describes the \emph{shearing wave}. From~\eref{orr:basic:shearingwavesolution} we can compute $u' = ik_y(t) \z' / k^2(t)$ and $v' = -ik_x \z' / k^2(t)$.

We next compute the net energy change of the mean flow due to a single wave that shears over and eventually dissipates.  We combine the time-dependent shearing wave solution with our previous energetics calculations.  We require the zonal average $\ol{u'v'}$, which is quadratic in wave fields.  The wave fields are ultimately real, and accounting for their complex representation, we have
	\begin{equation}
		\ol{u'v'} \to \frac{1}{2} \Re\bigl( u' v'^* \bigr) = -\frac{1}{2} \frac{k_x k_y(t)}{k(t)^4} |Z(t)|^2.\label{eq:uv_onewave}
	\end{equation}
The change in $E_{\mathrm{ZF}}$ is obtained by integrating~\eref{orr:zf_energetics_1} over the lifetime of the shearing wave:
	\begin{equation}
		\D E_{{\rm ZF}} = \int_0^\infty dt \d{E_{{\rm ZF}}}{t} = -\frac{1}{2} S \int_0^\infty dt \frac{k_x k_y(t)}{k(t)^4} |Z(t)|^2.\label{eq:DEzf}
	\end{equation}
We have used that $\partial_y \ol{u} = S$ (independent of $y$) and that the stress $\ol{u' v'}$ for an individual wave is also independent of $y$, and therefore the average over $y$ does nothing.

Equation~\eqref{eq:DEzf} shows that waves starting off with $k_y/k_x>0$ (quadrants I and~III in the $\v{k}$ plane) will take energy from the zonal flow, while waves starting off with $k_y/k_x<0$ (quadrants II and~IV) will give energy to the zonal flow. 
The simplest form of the  Kelvin--Orr shearing wave dynamics for growth of the shear flow arises from considering \emph{two} waves at the same amplitude, with initial wavevectors $(k_x, k_{y0})$ and $(-k_x, k_{y0})$. In isolation, one of the waves would grow in expense of the mean flow while the other would decay and give energy to the mean flow.  The two waves must be considered together because the net leading order contribution to $\D E_{{\rm ZF}}$ cancels out.  We ignore interactions between waves, meaning that in the computation of the stress $\ol{u'v'}$, we ignore cross terms.

The total energy change of the zonal flow $\D E_{{\rm ZF}}$ is the sum of that of the two waves individually, given by
	\begin{equation}
		\D E_{{\rm ZF}} = \frac{S}{2} \int_0^\infty dt \left[ \frac{k_x k_{y+}(t)}{k_+(t)^4} |Z_+(t)|^2 - \frac{k_x k_{y-}(t)}{k_-(t)^4} |Z_-(t)|^2 \right].
	\end{equation}
Here, a term with subscript $\pm$ stems from the wave with initial wavevector $(\mp k_x, k_{y0})$, where $k_{y\pm}(t) \equiv k_{y0} \pm S k_x t$.  We take the initial amplitudes $Z_+(0) = Z_-(0) = Z_0$.  From \eref{orr:basic:shearingwavesolution}, we have $|Z_\pm(t)|^2 = |Z_0|^2 e^{-2\n \int_0^t d\t\, k_\pm(\t)^2}$.  Assuming $k_x S t / k_{y0} \ll 1$, expanding to leading order in $S$, and dropping the $0$ subscript on $k_{y0}$, we obtain
	\begin{equation}
		\D E_{{\rm ZF}} = \frac{S^2 k_x^2 |Z_0|^2}{4 \n^2 k^4} \frac{k_x^2 - 5 k_y^2}{k^6}.
	\end{equation}
One immediate conclusion is that a pair of waves with wavevectors at a shallow enough angle to the $k_x$-axis tends to contribute energy to the mean flow, reinforcing it.  The critical angle is given by $\tan(\p_{\text{crit}}) = 1/\sqrt{5}$, or $\p_{\text{crit}} \approx 24^\circ$.\footnote{We note that modifying viscosity to instead be hyperviscosity of the form $\nu k^{2p}$ changes the critical angle to $\tan^{-1}\bigl[(3+2p)^{-1/2}\bigr]$.}  A pair of waves with an angle greater than $\p_{\text{crit}}$ draws energy from the mean flow, diminishing it.

We briefly comment on the use of periodic boundary conditions, infinite plane waves, and linear shear, which are mathematically convenient but could potentially raise some concern because of possible inconsistencies or physical subtleties.  Within the literature, others have explored the use of more realistic profiles for the shear flow and perturbations, such as using wavepackets rather than infinite plane waves.

These more realistic profiles have not been found to fundamentally alter the direction of energetic transfer from those in simpler calculations.  For instance, in a calculation involving perturbation growth in a finite-amplitude linear shear flow, \citet{farrell:1987} used localized perturbation wavepackets and showed that similar conclusions about energetic changes are obtained as when infinite plane waves are used.  Another calculation, more relevant to the present study as it is concerned with the growth of the mean flow, uses localized wavepackets and periodic, rather than linear, shear flow \citep{parker:2014-book}.  That calculation found energy transfer to the shear flow, just as is found here.


\subsection{With $\b$ effect and magnetic fields}
\label{subsec:Orr_betaB}
We extend now the analysis of the Kelvin--Orr shearing wave to include magnetic fields.  Again, we consider the energetics of the zonally averaged flow.  We neglect $\bar{A}$, which is justified by the later numerical findings in Section~\ref{sec:ZI}.

Returning to~\eref{eq:ubar}, retaining the magnetic fluctuations, and performing similar steps as in the previous subsection, we find the energetics of the mean zonal flow are now given by
	\begin{equation}
		\d{E_{{\rm ZF}}}{t} = \frac{1}{L_y} \int_0^{L_y} dy \, \Bigl( \ol{u' v'} \partial_y \bar{u} - \ol{B'_x B'_y} \partial_y \bar{u} \Bigr).
	\end{equation}
We also need the generalization of the shearing wave that includes magnetic fields.  With $\ol{A}=0$ and $\ubar = Sy$, the linearized, non-forced equations for the perturbations $\z'$ and $A'$ are
	\begin{subequations}
	\begin{align}
		(\partial_t + Sy \partial_x) \z' + \b \partial_x \psi' &= - B_0 \partial_x \nabla^2 A' + \n \nabla^2 \z' , \\
		(\partial_t + S y \partial_x) A' &= - B_0 \partial_x \psi' + \eta \nabla^2 A'.
	\end{align}
	\end{subequations}
Assuming $\z'(\v{x}, t) = Z(t) e^{i\v{k}(t) \cdot \v{x}}$ and $A'(\v{x}, t) = a(t) e^{i \v{k}(t) \cdot \v{x}}$, we find the same shearing dependence for~$\v{k}(t)$ as before (cf.~\eref{eq:shearwavenumbers}).  Then, we have\begin{subequations}
	\begin{align}
		\frac{dZ}{dt} &= \frac{i k_x \b}{k(t)^2} Z + i k_x B_0 k(t)^2 a - \n k(t)^2 Z ,\\
		\frac{da}{dt} &= \frac{i k_x B_0}{k(t)^2} Z - \eta k(t)^2 a .
	\end{align}
	\end{subequations}

If $k^2$ did not depend on time, then these equations would be exactly the linearized equations without mean flow and would give rise to the fast and slow waves with frequencies $\wf$, $\ws$.  In that case, the solution for any initial condition could be decomposed into the fast and slow eigenmode components.  In particular, if the linear combination $Z(0)$ and $a(0)$ start off exactly in the fast eigenmode, then the time dependence of $Z(t)$ and $a(t)$ is given by $\exp( -i \w_f t)$, where the imaginary part of $\w_f$ determines the damping rate.

The shear flow complicates matters because $k^2$ now changes with time.  However, when the shear is small, such that $k_x St/k_{y0} \ll 1$, $k^2$ remains nearly constant up through the decay time of the wave.  Hence, the constant-$k^2$ solution of the previous paragraph is mostly retained.  We expand $k^2$ to leading order in $S$.  If a wave starts as an eigenmode, it will stay in that eigenmode to lowest order; the solution for $Z$ is then given by
	\begin{equation}
		Z(t) = Z_0 e^{-i \th(t)} \exp \left[\int_0^t d\t\, \Im \w(\t) \right],\label{eq:Zsol}
	\end{equation}
where $\th(t) \equiv \int_0^t d\t\, \Re \w(\t)$ is some phase. An expression similar to \eref{eq:Zsol} also holds for $a(t)$.

We now restrict ourselves to the parameter regime where $\n k^2$, $\eta k^2$, $\w_A \ll \w_R$.  The fast and slow frequencies $\w_f$ and $\w_s$  simplify to the expressions in \eref{eqs:fastslow_easyregime}.  In this limit, $\Im(\w_f) = -\n k^2$ and $\Im(\w_s) = -\eta k^2$, and the wave damping behaves purely diffusively.  When starting in the fast eigenmode, the solution for small shear is
	\begin{subequations}
	\label{eq:shearingwaveB}
	\begin{align}
		Z_f(t) &= Z_0 e^{-i \th_f(t)} \exp \left[-\n \int_0^t d\t\, k(\t)^2 \right], \\
		a_f(t) &= A_0 e^{-i \th_f(t)} \exp \left[-\n \int_0^t d\t\, k(\t)^2 \right].
	\end{align}
	\end{subequations}
The initial amplitudes $Z_0$ and $A_0$ are related by the eigenmode relation~\eref{eigenmode_relations_easyregime}.  A similar expression exists for the slow wave, with~$\n$ replaced by~$\eta$.

We use the shearing wave solution in~\eref{eq:shearingwaveB} to compute the energetic changes of the mean flow.  For a single wave, the Reynolds stress $\ol{u'v'}$ is given by~\eref{eq:uv_onewave}; similarly the Maxwell stress is given by
	\begin{equation}
		\ol{B_x' B_y'} \to \frac{1}{2} \Re\bigl( B_x' B_y'^* \bigr) = -\frac{1}{2} k_x k_y(t) |a(t)|^2.
	\end{equation}

Integrating over the lifetime of the wave, the net energy change in the mean flow due to a single wave shearing over is then
	\begin{equation}
		\D E_{{\rm ZF}} = -\frac{1}{2} Sk_x \int_0^\infty dt \left[ \frac{k_y(t)}{k(t)^4} |Z(t)|^2 - k_y(t) |a(t)|^2 \right].\label{eq:DEzf_magnetic}
	\end{equation}
We consider the effect of two (noninteracting) waves, with initial wavevectors $(k_x, k_{y0})$ and $(-k_x, k_{y0})$.  The procedure is much the same as in Section~\ref{subsec:Orr}.  Expanding to leading order in $S$, we obtain
	\begin{equation}
		(\D E_{{\rm ZF}})_{f} = \frac{S^2 k_x^2}{4 \n^2 k^4} \bigg( \underbrace{\frac{k_x^2 - 5 k_y^2}{k^{6}} |Z_0|^2 - \frac{k_x^2 - k_y^2}{k^2}|A_0|^2}_{\equiv J} \bigg) .
		\label{orr:fastwave_energy}
	\end{equation}
The corresponding expression for $(\D E_{{\rm ZF}})_{s}$ is identical with $\n$ replaced by~$\eta$.

Expression~\eqref{orr:fastwave_energy} generalizes the energy transfer to a weak mean flow due to Kelvin--Orr shearing wave dynamics to include magnetic fields. It is a major result of this paper. The term proportional to $|Z_0|^2$ stems from the Reynolds stress while the term proportional to $|A_0|^2$ comes from the Maxwell stress.  We note that no explicit dependence on $\b$ or $B_0$ has yet appeared in~$\D E_{{\rm ZF}}$.  Both~$\b$ and $B_0$ have only an indirect effect on the size of the perturbations $Z_0$ and $A_0$.

Focusing on the wavevector dependence, we examine the quantity $J$ inside the parentheses in~\eref{orr:fastwave_energy}, which $(\D E_{{\rm ZF}})_{f,s}$ is proportional to.  Substituting the energy-normalized eigenfunctions from \eref{eigenmode_relations_easyregime} and letting $k_x = k \cos \p$ and $k_y = k \sin \p$, we obtain
	\begin{subequations}
	\begin{align}
		J_f &= (\cos^2 \p - 5 \sin^2 \p) - \frac{\w_A^2}{\w_R^2} ( \cos^2 \p - \sin^2 \p) , \label{eq:orr:Jf} \\
		J_s &= \frac{\w_A^2}{\w_R^2} (\cos^2 \p - 5 \sin^2 \p) - ( \cos^2 \p - \sin^2 \p) ,
	\end{align}
	\end{subequations}
for the fast and slow wave, respectively.

\begin{figure}
		\includegraphics[width=\columnwidth]{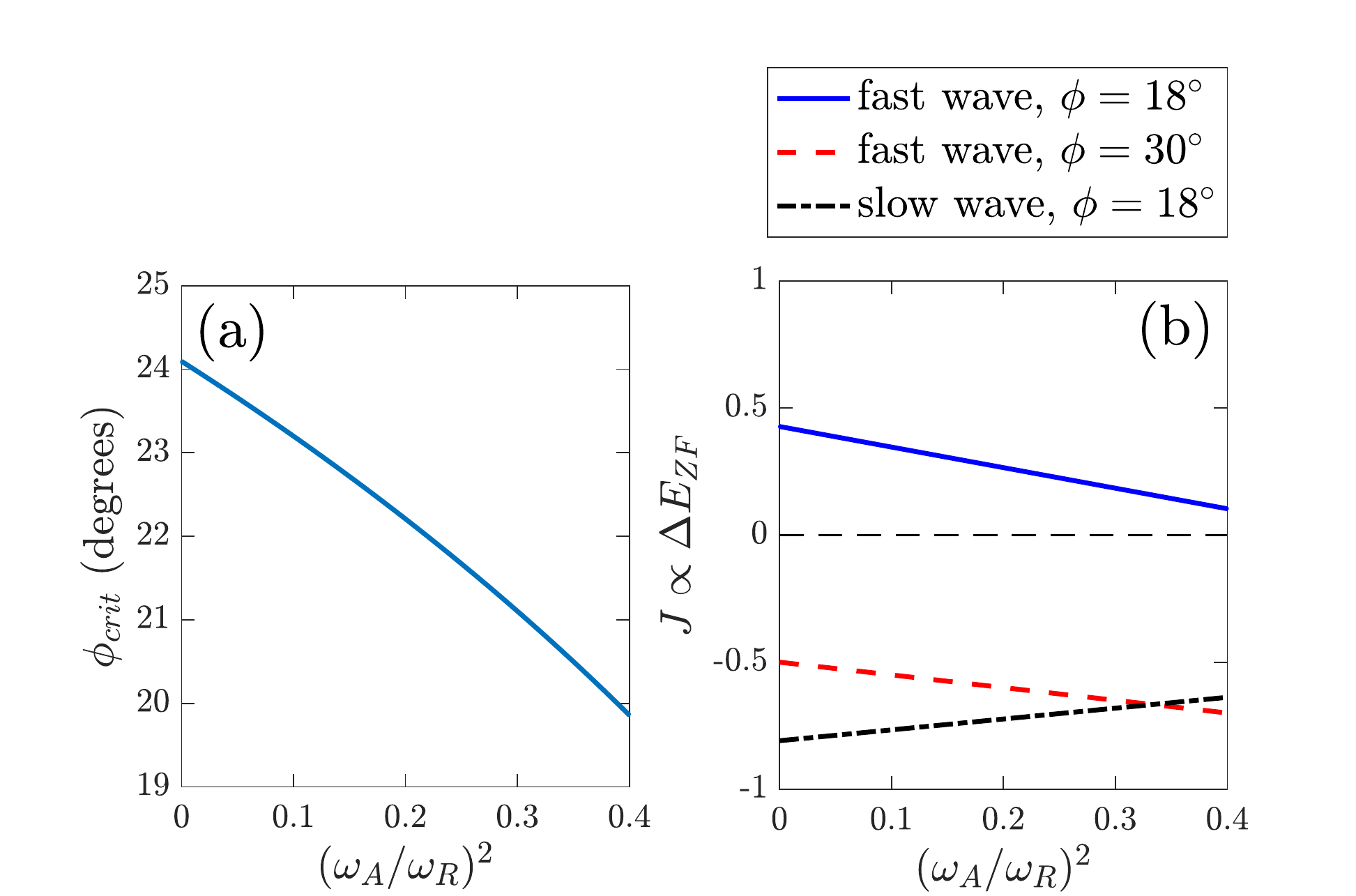}
		\caption{(a) Critical angle $\p_{crit}$ below which the fast wave contributes to driving a mean zonal flow perturbation, as a function of normalized background magnetic field, $\w_A / \w_R$.  An increasing magnetic field decreases the critical angle, allowing fewer wavevectors to drive mean flow growth.  (b) The quantity~$J$, which is proportional to the change in energy of the mean flow, as a function of normalized magnetic field, at fixed angle $\p = \tan^{-1} (k_y / k_x)$.  For $\p < \p_{crit}$, as the magnetic field increases, for the fast wave, $J_f$ decreases in magnitude, and for the slow wave, $J_s$ somewhat decreases in magnitude.}
		\label{fig:orr_mech}
	\end{figure}

We make several observations.  First, for the fast wave, $J_f = 0$ determines the critical angle $\p_{crit}$ that separates the waves that drive the mean flow from those that suppress it.  As mentioned before, without magnetic field, $\p_{crit}\approx 24^\circ$. 
However, from \eref{eq:orr:Jf}, we see that turning on the magnetic field causes the second term to become nonzero.  Increasing the magnetic field reduces the critical angle, implying that now a smaller subset of fast-wave perturbations can contribute positively toward the growth of the shear flow.  Figure~\ref{fig:orr_mech}(a) shows how the critical angle varies with the background magnetic field~$B_0$.  Waves with $\p < \p_{crit}$ are of primary interest because these fast waves contribute positively to $\D E_{{\rm ZF}}$, potentially driving strong growth of the mean flow.  Second, $J_s$, like~$J_f$, can be of either sign.  However, for those waves with $\p < \p_{crit}$, the slow wave opposes the mean flow, i.e., $J_s < 0 $.  Third, for $\p < \p_{crit}$, the magnitude of~$J_f$ decreases as the magnetic field increases.  The magnitude of $J_s$ also somewhat decreases (see Figure~\ref{fig:orr_mech}(b)).

The Kelvin--Orr shearing wave calculation does not capture the relative fraction of energy that resides in magnetic fluctuations compared with hydrodynamic fluctuations.  Rather, the strength of these fluctuations, $Z_0$ and $A_0$, are taken here as given.  In the parameter regime we have examined, magnetic fluctuations reside primarily in the slow wave and hydrodynamic fluctuations reside in the fast wave.  Because $Z_0$ and $A_0$ are exogenous to this calculation, the use of energy-normalized eigenfunctions eases the interpretation of the physics by separating the effect of the wave from the amount of energy contained in each wave.  Intuitively, and as we shall see later, as~$B_0$ increases, more energy resides in the magnetic fluctuations and the slow wave more strongly suppresses the growth of zonal flow.  

We have assumed an initial condition that starts off as a pure fast or slow wave, and calculated the effects of the two waves separately.  Mathematically, this is equivalent to neglecting cross terms in the Reynolds and Maxwell stresses, which are quadratically nonlinear.  From a physical point of view, this amounts to an assumption that the interaction between waves is negligible.

To summarize this section, we have generalized the Kelvin--Orr shearing wave for a weak shear flow to include magnetic fields.  We obtained \eref{orr:fastwave_energy}, one of the major results of this article, which describes a mean shear flow's energetic change due to a pair of shearing waves.  Our calculation shows that magnetic fluctuations, through the slow magneto-Rossby wave, will oppose the growth of a mean shear flow.  An additional effect is that a stronger $B_0$ also reduces the fast wave's contribution to driving a mean flow.  We shall see later that the former is the dominant effect (see ~Figure~\ref{fig:stresses_vs_eta}(b) and surrounding discussion).

The Kelvin--Orr calculation is not a complete description because it does not close the loop and say how the zonal flow dynamically evolves.  Furthermore, the computation is limited to long-wavelength shear flows.  It also does not provide a growth rate.  But it does give a clear physical picture of the effect of a weak shear flow on fluctuations, and shows, unambiguously, that a magnetic field opposes the growth of zonal flows.  This simple calculation also quantitatively predicts which wavevectors contribute to driving or suppressing zonation.

The next section includes a more detailed and elaborate computation that is both dynamically consistent and also is not limited to long-wavelength mean flows.  We shall see that the key conclusions of the wavenumber dependence of the Reynolds and Maxwell stress found in this simple Kelvin--Orr calculation [\eref{orr:fastwave_energy}] are recovered from the more consistent calculation of the next section, in the appropriate asymptotic limit.

\section{Zonostrophic instability with magnetic field}
\label{sec:ZI}

The CE2 dynamical system in Section~\ref{sec:CE2formulation} exhibits a homogeneous equilibrium that consists of zero mean fields, $\ol{u}=0$ and $\ol{A}=0$, and eddy covariances that are homogeneous in both spatial directions, e.g.,~$W(\v{x}_a, \v{x}_b) = W^H(\v{x}_a-\v{x}_b)$, etc. This equilibrium can become unstable to zonal jets in what is known as zonostrophic instability~(ZI).

We analyze here the zonostrophic instability of~\eref{eq:EOM}. That is, we ask if perturbations about the homogeneous equilibrium, $\delta\bar{u}$, $\delta\bar{A}$, along with eddy covariance perturbations, e.g., $W=W^H + \delta W$, lead to exponential growth. The mean field perturbations are written as, e.g., $\delta\bar{u}=c_u\,e^{\l t} e^{i q y}$. If there exists $\lambda$ with positive real part we say that the homogeneous equilibrium is unstable and leads to mean flow growth at wavenumber $q$.
The techniques for the stability calculations are standard; the reader is referred, e.g.,  to the work by~\citet{srinivasan:2012}, in which the same type of calculation was carried out for an unmagnetized barotropic fluid. We provide the backbone of the calculation in the \hyperref[app:zonostrophic]{Appendix}.

\subsection{Zonostrophic instability results}
\label{sec:ZIexample}

We present results from the ZI analysis. We consider a domain of size $2\pi\times2\pi$, use parameter values $\beta=2$, $\nu=\eta=10^{-4}$, and take isotropic forcing centered about a total wavenumber~$k_f$. That is:
\begin{equation}
	\hat{Q}_{\v{k}} = Q_0\, e^{-(k-k_f)^2/(2\,\delta k_f^2)} ,
\end{equation}
where
\begin{equation}
	Q_0 = 5\times10^{-5} ,\ k_f=12,\ \text{and}\ \delta k_f =1.5 .
\end{equation}
This forcing injects energy into hydrodynamic fluctuations at a rate $\e = \sum_{\v{k}} \hat{Q}_{\v{k}}/(2k^2) = 4.81\times 10^{-5}$. The forcing introduces a length scale~$k_f^{-1}$ and a time-scale~$(\epsilon k_f^2)^{-1/3}$. 

\begin{figure}
\includegraphics[width=\columnwidth]{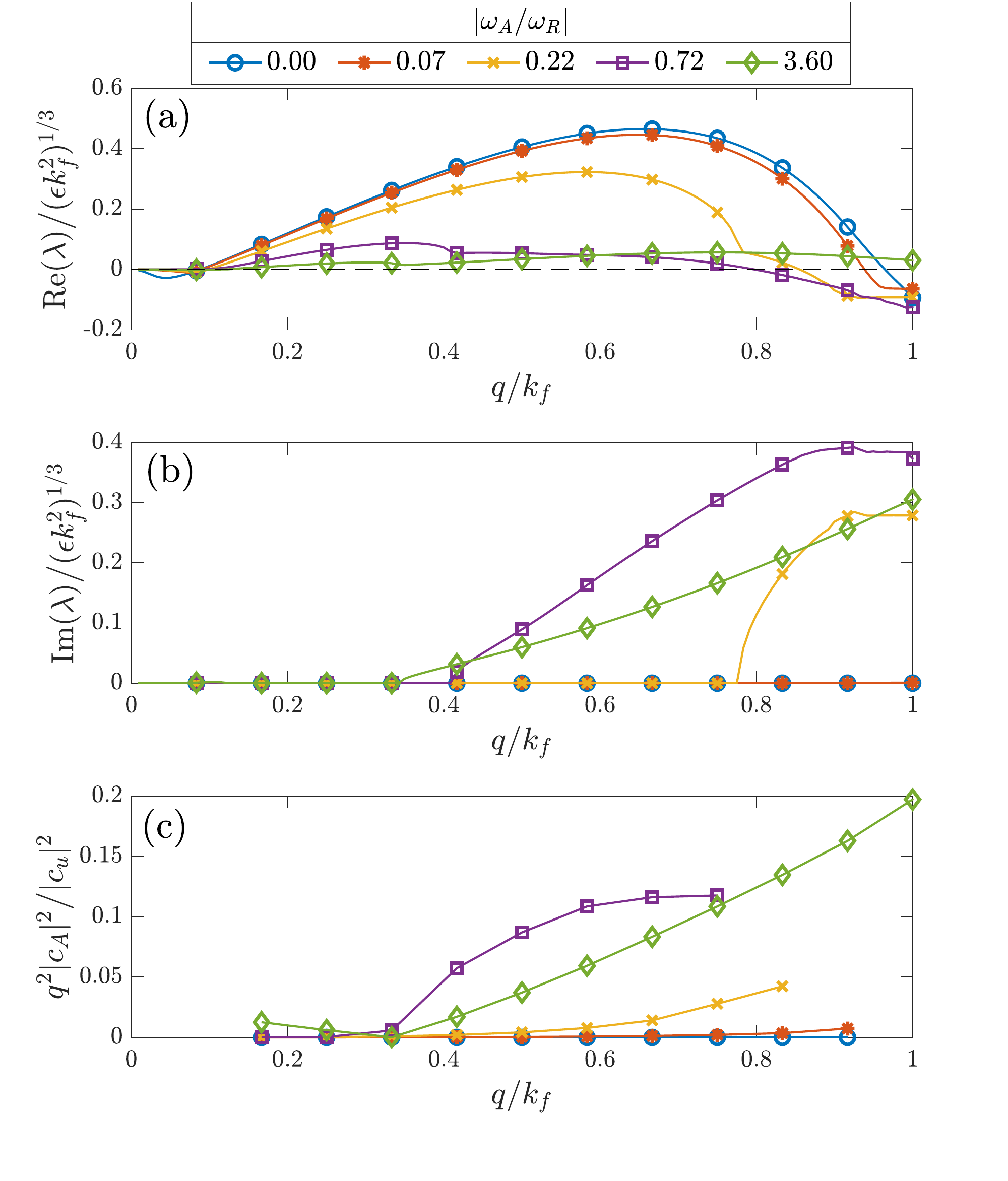}
\vspace{-2.5em}
\caption{Most unstable ZI eigenvalue $\l$ as a function of the mean flow wavenumber $q$ for the case discussed in Section~\ref{sec:ZIexample} (panels (a),~(b)). (Dots mark the mean-field wavenumbers that fit in our domain.) For the unstable cases, panel~(c) shows the ratio of the magnetic energy to the zonal flow energy $q^2|c_A|^2\big/|c_u|^2$. Magnetic energy is much less than the zonal flow energy; the energy ratio goes up to 0.2 but that happens for $|\wA/\wR|\ge3.60$ for which $\l$ come with weak growth rates and are also complex. \label{fig:la_q_fixB0}}
\end{figure}

\begin{figure}
	\includegraphics[width=\columnwidth]{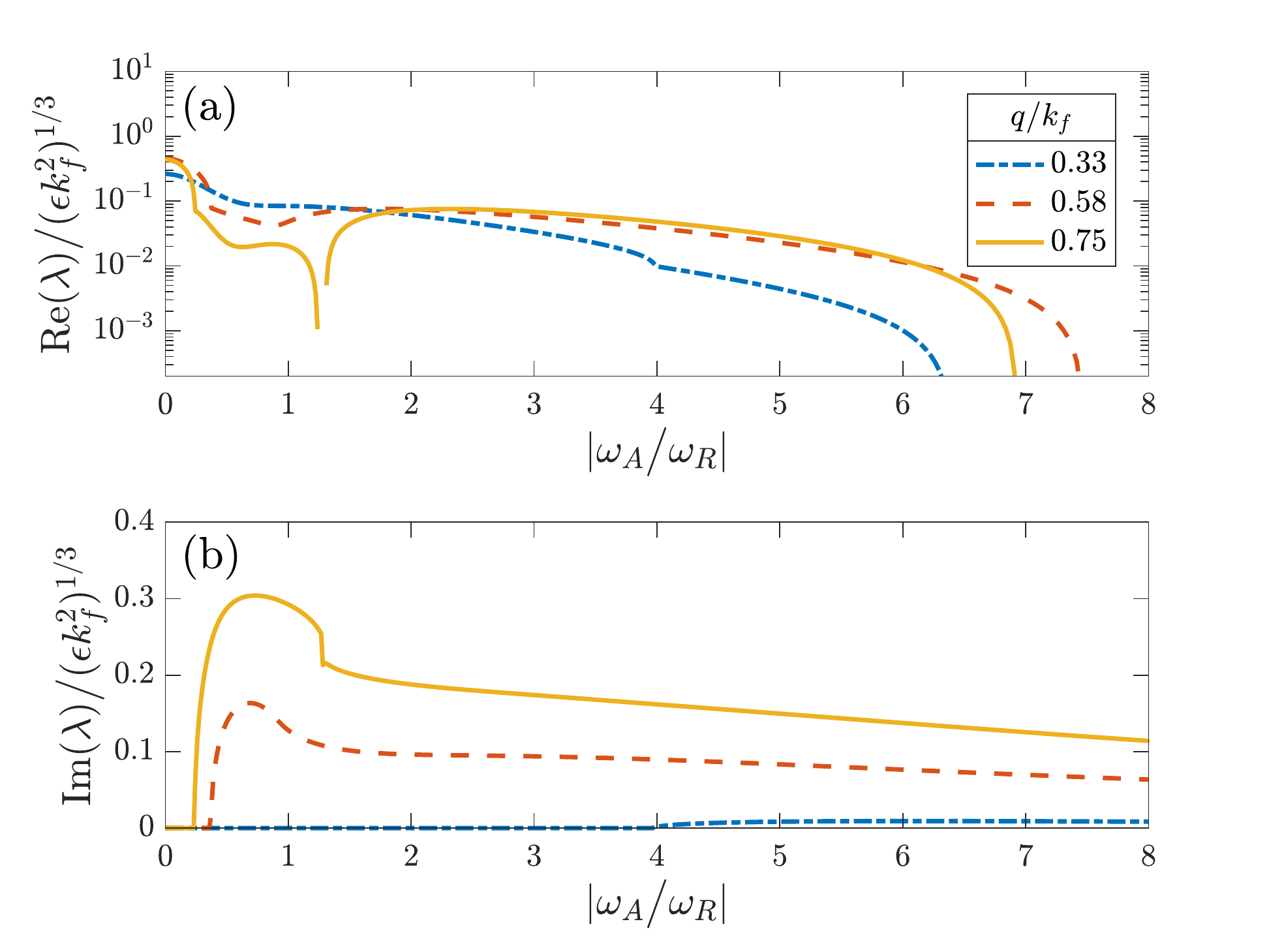}
	\vspace{-1.5em}
	\caption{Most unstable ZI eigenvalue $\l$ as a function of the background magnetic field $B_0$ (all other parameters held fixed) for the case discussed in Section~\ref{sec:ZIexample}.  When $|\w_A / \w_R| \lesssim 0.25$, the growth is strongest (largest real part), and the eigenvalue is real.  For larger $|\w_A / \w_R|$, not only does the growth weaken considerably, but also the eigenvalue becomes complex.\label{fig:lam_B0_fixq}}
\end{figure}

\begin{figure*}[t]
	\centering
	\includegraphics[width=0.7\textwidth]{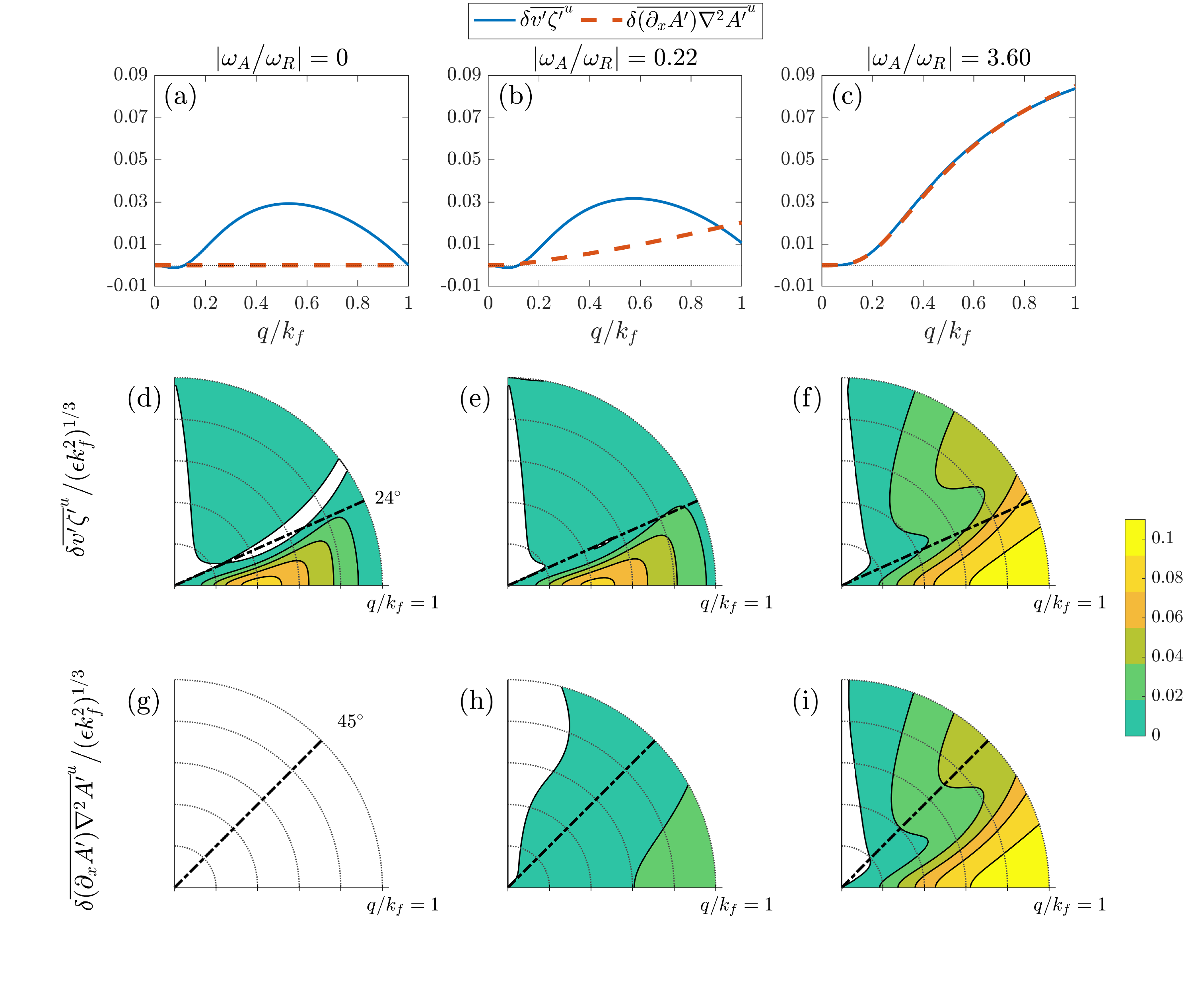}
	\caption{Comparison of the Reynolds and Maxwell stresses for marginally stable ($\l = 0$) eigenmodes.  Panels (a)--(c) show the total Reynolds stress (solid) and Maxwell stress (dashed) for three values of the background magnetic field $B_0$.  The rest of the panels show the spectral decomposition of these total stresses into their contributions from individual eddy wavevectors.  The spectral decomposition of the Reynolds stress is shown in (d)--(f) and the Maxwell stress in (g)--(i).  Stresses are shown on a $(q,\phi)$ polar grid: values shown correspond to the net contribution to the stresses from the four modes $\v{k} = k_f \times \{(\cos\p, \sin\p), (-\cos\p, \sin\p), (-\cos\p, -\sin\p), (\cos\p, -\sin\p)\}$ on a mean zonal flow perturbation with wavenumber $q$. For the Reynolds stress, positive values (yellow or green) reinforce the zonal flow and negative values (white) oppose it.  For the Maxwell stress, positive values oppose the zonal flow and negative values reinforce it.  The stresses were computed using Eqs.~\eqref{eq:dvz}--\eqref{eq:dAlapA} at the marginal point for ZI ($\l=0$). Contour levels start at 0 and increase by 0.02; dash--dotted lines mark the critical angles $\phi_{crit} \approx 24^\circ,~45^\circ$ (see Section~\ref{sec:Orr}).  At high enough $B_0$, the Maxwell stresses become identical with the Reynolds stresses and thus ZI is suppressed. \label{fig:stresses_qphigrid}}
\end{figure*}

For each $q$, there are multiple eigenmodes, each with its own ZI eigenvalue $\lambda$. Figure~\ref{fig:la_q_fixB0} shows the eigenvalue with maximum growth rate as a function of the mean flow wavenumber~$q$ for various values of the strength of the background magnetic field $B_0$ (normalized as $|\w_A / \w_R|$). As $B_0$ increases, the ZI is inhibited. This inhibition is also seen in Figure~\ref{fig:lam_B0_fixq} in which the eigenvalue $\l$ is shown as a function of the magnetic field strength for fixed mean-field wavenumber $q$.

When there is instability, the mean-flow components of the eigenfunction consists primarily of mean zonal jet $\delta\bar{u}$ rather than mean magnetic field $\delta\bar{A}$; see Figure~\ref{fig:la_q_fixB0}(c). That the mean flow eigenfunction is dominated by $\delta\bar{u}$ is a general characteristic of the ZI of \eref{eq:EOM}, at least in all parameter ranges we have explored.  The smallness of the mean magnetic component compared to the mean flow justifies our choice in the Kelvin--Orr calculation (section \ref{subsec:Orr_betaB})  to use only a mean shear flow and to neglect a mean sheared magnetic field.

When the ZI is robustly strong---typically at low values of the magnetic field, $|\w_A / \w_R| \lesssim 0.25$---the eigenvalue is typically real.  As the magnetic field becomes stronger, not only does the growth rate drop considerably, but also the eigenvalue becomes complex; this is seen in both Figures~\ref{fig:la_q_fixB0} and~\ref{fig:lam_B0_fixq}.  While our ZI calculation is only linear and does not predict the final nonlinearly saturated state, the physics of a stationary (real eigenvalue) and translating (complex eigenvalue) mode can be quite different, and it is useful to distinguish between these cases.  For instance, it is possible that the growing mode with real eigenvalue saturates into stationary zonal flows, while the mode with complex eigenvalue does not.

We can gain insight into how the ZI is inhibited by examining the Reynolds and Maxwell stresses for the eigenmodes.\footnote{We reiterate that we are using the term Reynolds stress as a shorthand, when we are actually referring to the divergence of the Reynolds stress.}  Recalling the zonally averaged momentum \eref{eq:ubar}, the Reynolds and Maxwell stresses are the fluctuation-driven terms that can drive or oppose the growth of the mean flow.

The perturbation equation for the mean flow eigenmode is described by
	\begin{equation}
	(\l + \nu q^2)\de\bar{u} - \de \overline{v'\z'} + \de \overline{(\partial_x A')\nabla^{2} A'} = 0,
	\end{equation}
which comes directly from \eref{eq:ubar}.  To a good approximation the above simplifies to
\begin{equation}
	\l + \nu q^2 - \de \overline{v'\z'}^u + \de \overline{(\partial_x A')\nabla^{2} A'}^u = 0.
	\end{equation}
Here, the $u$ superscript refers to the parts of the stresses associated with the perturbation mean flow $\de \bar{u}$, neglecting the contribution associated with the perturbation mean magnetic field $\de \bar{A}$.  This decomposition of the stresses into components associated with~$\de \bar{u}$ and~$\de \bar{A}$ emerges from the instability calculation detailed in the \hyperref[app:zonostrophic]{Appendix}.  Because the mean magnetic component of the eigenfunction is small, $\de \overline{v'\z'} \approx e^{iqy}\,\de \overline{v'\z'}^u$ and $\de \overline{(\partial_x A')\nabla^{2} A'} \approx e^{iqy}\, \de\overline{(\partial_x A')\nabla^{2} A'}^u$.

Figure~\ref{fig:stresses_qphigrid} shows the fluctuation stresses $\delta\overline{v'\z'}^u$ and $\delta\overline{(\partial_x A')\nabla^{2} A'}^u$.  Panels (a)--(c) show the Reynolds stress and Maxwell stress for a marginally stable eigenmode $\l=0$ at three different values of the magnetic field.  In this figure, positive values of the Reynolds stress \emph{reinforce} the zonal flow, while positive values of the Maxwell stress \emph{oppose} the zonal flow.  At zero magnetic field (panel (a)), the Maxwell stress is zero and the Reynolds stress drives growth.  At moderate magnetic field (panel (b)), the Maxwell stress is nonzero and opposes the zonal flow, but it does not have a significant effect because it is still considerably less than the Reynolds stress.  At a large magnetic field (panel (c)), the Maxwell stress has grown such that it is almost exactly equal to the Reynolds stress.  The Maxwell stress completely counteracts the driving effect of the Reynolds stress. 

It is also possible to take a closer look and examine the spectral decomposition of the Reynolds and Maxwell stresses.  Considering marginally stable modes has been a useful way in earlier studies of ZI in unmagnetized fluids to understand which of the spectral components of the forcing contribute to ZI \citep{bakas:2013-jas,bakas:2015}.  Using analytic formulas derived in the course of the ZI calculation, we can extract the contribution of individual Fourier modes to the stresses.  The procedure to obtain these analytic formulas is described in the \hyperref[app:zonostrophic]{Appendix}, but the formulas themselves are not written explicitly because they are extremely complicated. 

We can thus determine which fluctuation wavevectors tend to contribute positively or negatively toward the Reynolds and Maxwell stress.  Figure~\ref{fig:stresses_qphigrid}(d)--(i) depict the spectral decomposition of marginally stable eigenmodes on a $(q, \phi)$ polar grid.  For example, for the case with $B_0=0$, panel (d) implies that when a mean-flow perturbation $\delta\bar{u}$ with wavenumber $q/k_f=0.4$ is introduced in the flow, the forcing components $\v{k}=(\pm k_f,0)$ will induce Reynolds stresses with $\delta\ol{v'\z'}^u\approx0.08 (\epsilon k_f^2)^{1/3}>0$ that tend to reinforce $\delta\bar{u}$, leading to instability.  

We can see that for small values of the background magnetic field, the contribution of each component of the forcing to the Reynolds stresses remains mostly unchanged.  In other words, panel (e) is mostly the same as panel (d).  On the other hand, panel (h) shows the spectral decomposition of the Maxwell stress at moderate magnetic field.  For high values of the magnetic field, comparison of panels (f) and (i) shows that the cancellation between Reynolds stresses and Maxwell stresses occurs at each individual wavevector.

At this point, we can make close connection with the Kelvin--Orr shearing wave calculation presented in Section \ref{subsec:Orr_betaB}.  The analytic formulas used for the spectral decompositions of the Reynolds and Maxwell stresses in Figure~\ref{fig:stresses_qphigrid} can be asymptotically expanded in a limit relevant to the Kelvin--Orr shearing wave.  The limit consistent with the Kelvin--Orr calculation is to take $\l \to 0$, small $B_0$, and small $q$.  In this limit, the leading order terms for the Reynolds and Maxwell stresses are 
	\begin{subequations}
	\begin{align}
		\delta\ol{v'\,\z'}^u &= \frac{2k_x^2 q^2 \left(k_x^2-5 k_y^2\right)}{\nu  k^8}\hat{W}^H_{\v{k}},\\
		\delta\ol{(\partial_x A')\nabla^{2} A'}^u & = \frac{2k_x^2 q^2 \left(k_x^2- k_y^2\right)}{\eta k^4}\hat{G}^H_{\v{k}}.
	\end{align}
	\end{subequations}
The parameter scalings for the Reynolds and Maxwell stresses are remarkably similar to those in \eref{orr:fastwave_energy}.  (Recall that the first term on the right-hand side of \eref{orr:fastwave_energy} arises from the Reynolds stress, and the second term from the Maxwell stress.)  In particular, the wavevector dependence that determines positive vs.\ negative contribution, $(k_x^2 - 5k_y^2)$ for the Reynolds stress and $(k_x^2 - k_y^2)$ for the Maxwell stress, is exactly the same in the asymptotic limit of the ZI calculation and in the Kelvin--Orr calculation.  The computer algebra system Mathematica was used both to derive the expressions for the stresses and to take the asymptotic limit.

Figure~\ref{fig:stresses_qphigrid} shows, roughly, how small $q$ must be (i.e., how long wavelength the zonal flow must be) for this asymptotic limit to be accurate.  For example, we see that for $q/k_f\lesssim 0.2$ the Reynolds stresses are positive only for $\phi<24^\circ$; similarly, the Maxwell stresses are positive for $\phi < 45^\circ$.  For $q/k_f\gtrsim 0.2$, the constant-angle boundary (dash--dotted line) between positive and negative stresses is no longer accurate.

\begin{figure}
	\includegraphics[width=\columnwidth]{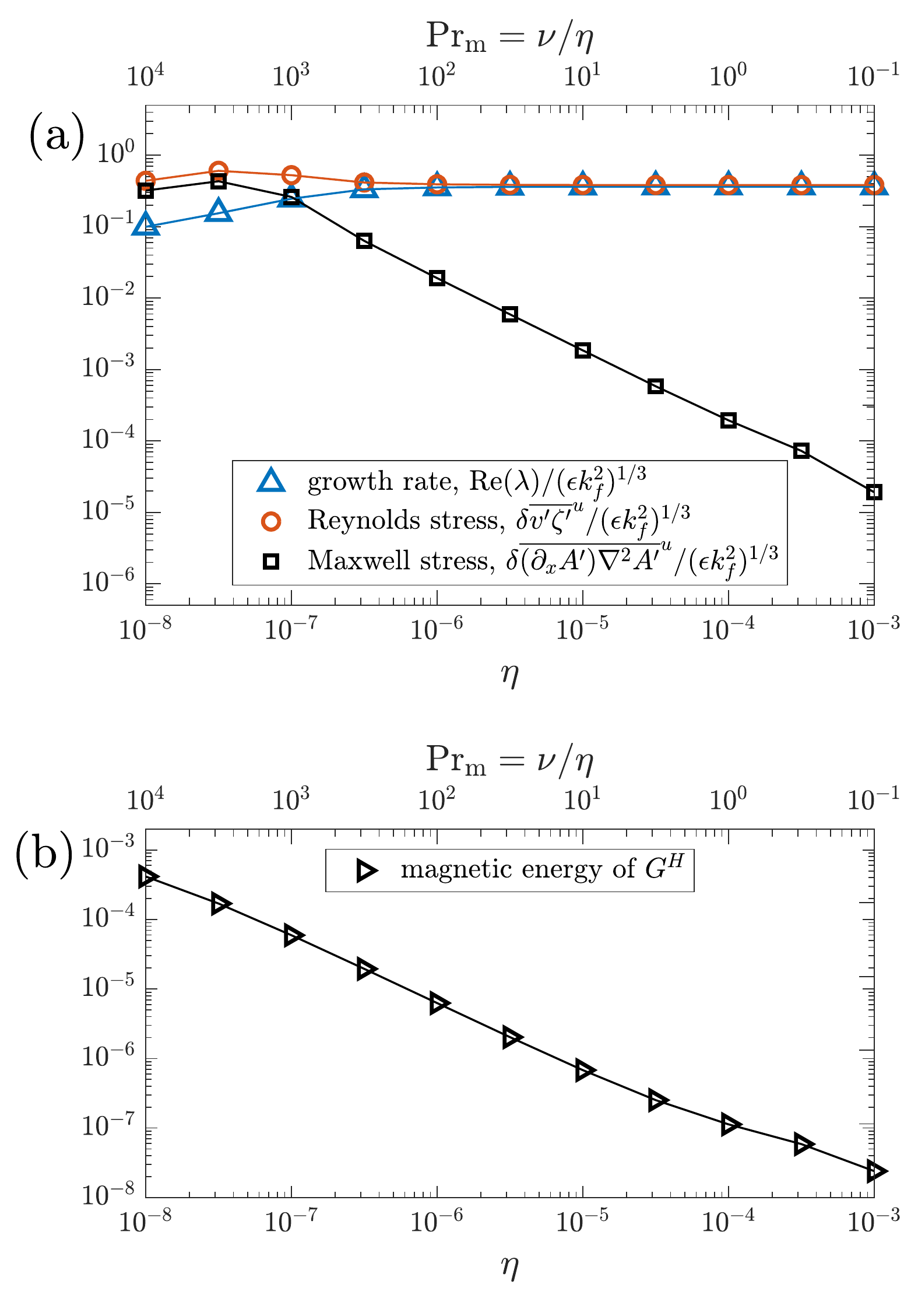}
	\caption{(a) Growth rate $\Re(\l)$ and Reynolds stress and Maxwell stress as functions of resistivity $\eta$.  A positive sign of the Maxwell stress opposes the growth of zonal flow.  As $\eta$ decreases, the Maxwell stress increases and the Reynolds stress is relatively unchanged, until the Maxwell stress becomes comparable to the Reynolds stress around $\eta = 10^{-7}$, and the growth rate of ZI drops sharply.  At $\eta = 10^{-8}$, the growth rate, Reynolds stress, and Maxwell stress are all complex, with an imaginary part on the same order of magnitude as the real part; only the real part is shown in the figure.  At the other values of $\eta$, these quantities are real.  (b) The magnetic energy of the magnetic fluctuation covariance $G^H$ increases as $\eta$ decreases. For both panels, the parameters used are $\nu = 10^{-4}$, $B_0 = 10^{-4}$, $\b=2$, $Q_0 = 4\times 10^{-5}$, $k_f=12$, and a fixed mode number of the zonal flow, $q=6$.  The ratio $\wA / \wR \approx 0.0072$.}
	\label{fig:stresses_vs_eta}
\end{figure}

Figure~\ref{fig:stresses_vs_eta} shows the balance between the Reynolds and Maxwell stresses as the resistivity $\eta$ changes.  As $\eta$ changes from large to small, the Maxwell stress grows larger (Figure~\ref{fig:stresses_vs_eta}(a)).  In Figure~\ref{fig:stresses_vs_eta}(b), we see that the Maxwell stress grows at the same rate as the overall level of magnetic fluctuations, as measured by the magnetic energy stored in the covariance $G^H$ of the CE2 homogeneous equilibrium.  At large $\eta$, for which magnetic fluctuations are suppressed, strong ZI occurs and the growth rate is about $0.4(\epsilon k_f^2)^{1/3}$.  As $\eta$ decreases and the level of magnetic fluctuations grows, eventually the Maxwell stress becomes comparable to the Reynolds stress, and the ZI is suppressed, with the growth rate weakening considerably.  The eigenvalue $\l$ and the stresses even become complex at $\eta = 10^{-8}$, whereas these quantities are real for larger $\eta$.

\begin{figure*}
	\centering
	\includegraphics[width=0.99\textwidth]{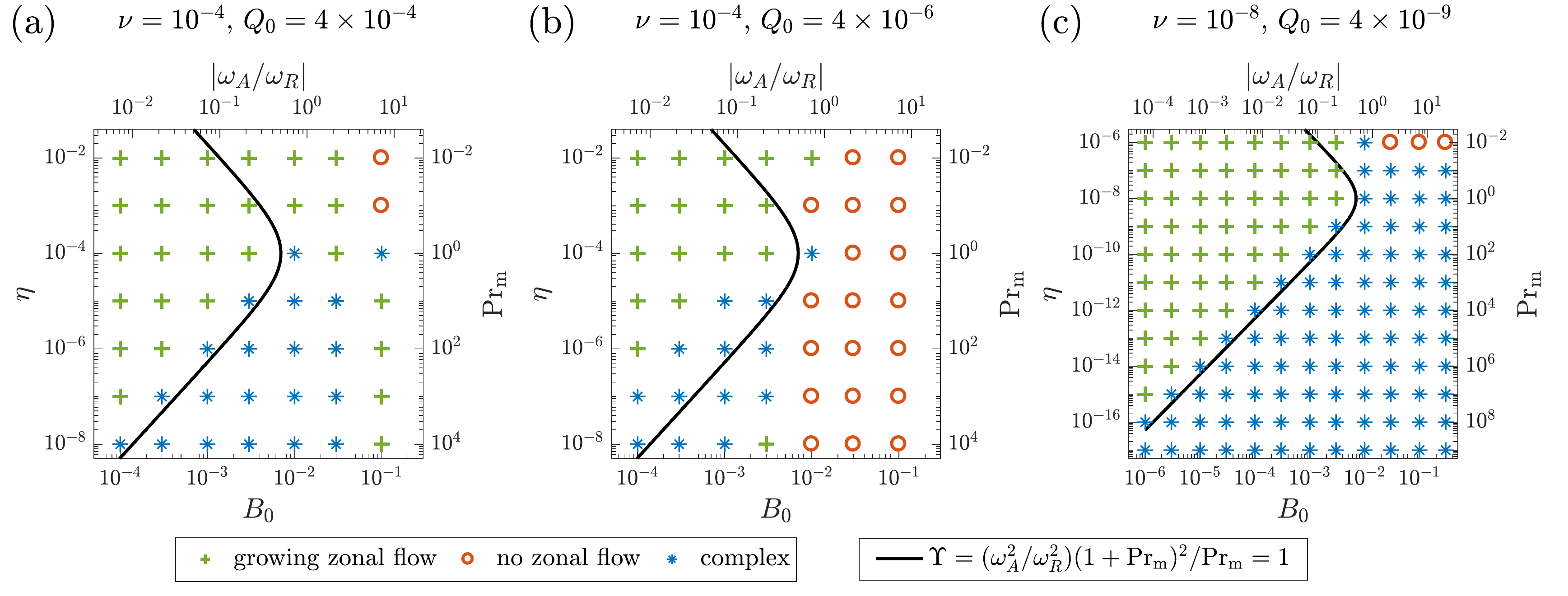}
	\caption{Behavior of ZI as $B_0$ and $\eta$ vary.  The three panels use different values of $\n$ and $Q_0$.  A single eigenmode wavenumber $q=6$ is used throughout.  For each value of $B_0$ and $\eta$, a marker depicts the type of behavior of the most unstable eigenmode: growing zonal flow (eigenvalue $\l$ is real and positive; plus signs), no zonal flow ($\l$ is real and negative; circles), or indeterminate ($\l$ is complex, often with positive real part; asterisks).  The parameters for panel (b) included a forcing amplitude two orders of magnitude weaker than that used in panel (a).  However, the boundary between growing zonal flow and the other behaviors is mostly unchanged between these two panels.  Also shown is the curve $\Y=1$ (see \eref{eq:upsilon}).  The bottom half of this curve, at which $\Prm \gg 1$, fits well the zonation boundary.  For $\Prm \gg 1$, $\Y = 1$ reduces to $\eta / B_0^2 = \text{constant}$.  In panel (c), another example is shown, with a much smaller value of $\n$.  In panel (a), there are some isolated examples of unstable modes at high $B_0$ and small $\eta$; it is not fully understood why these appear.\label{fig:eta_B0}}
\end{figure*}

Figures~\ref{fig:eta_B0}(a)--(c) show the behavior of ZI on an $(\eta, B_0)$ grid.  For each parameter value, a marker depicts whether the homogeneous equilibrium leads to growing, stationary ZF (ZI eigenvalue $\l$ is real and positive, plus~$+$ signs), no growing ZF ($\l$ is real and negative, circles~$\circ$), or something indeterminate ($\l$ is complex, often with positive real part, asterisks~$\ast$).   For these plots, only $\eta$ and $B_0$ change while all other parameters are kept the same.  Only a single ZF wavenumber $q=6$ is used, which is typically close to the most unstable wavenumber.  Figures~\ref{fig:eta_B0}(a)~and~(b) use the same parameters except the amplitude of the input forcing $Q_0$ is varied.  Figure~\ref{fig:eta_B0}(c) uses a different value of $\nu$.

\begin{figure}
	\centering
	\includegraphics[width=.8\columnwidth]{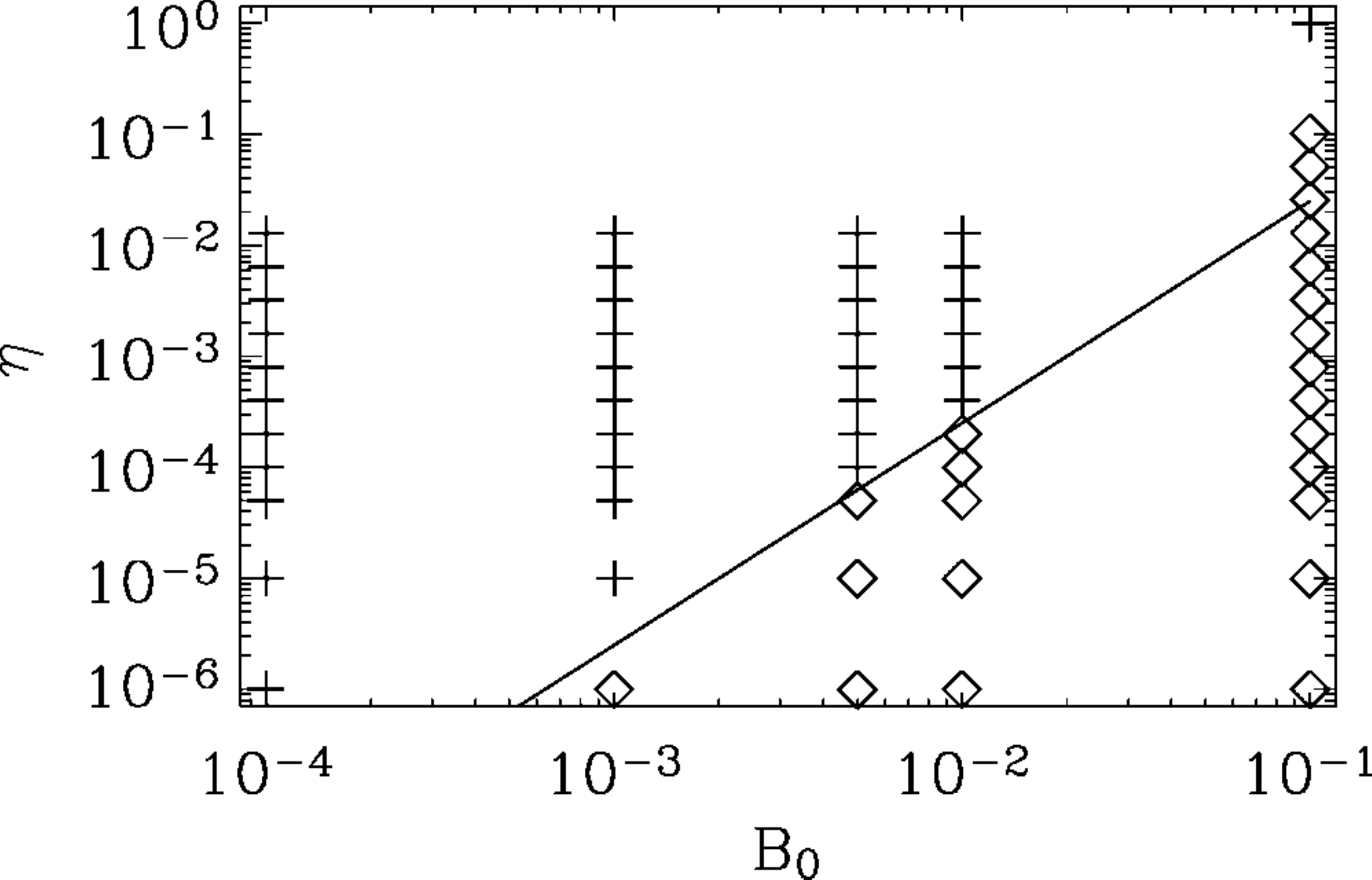}
	\caption{Nonlinear solutions of~\eref{eq:EOM} by \citet{tobias:2007}. Plus signs (+) denote cases with zonal jets are present; diamonds ($\diamond$) denote cases where zonal jets are inhibited. (Figure reproduced from the paper by \citet{tobias:2007}; copyright ApJ, 2007.)}
	\label{fig:etaB0_Tobias}
\end{figure}

Up to some maximum $B_0$, the boundary in $(\eta, B_0)$ space between the growing, stationary zonal flow and the other behaviors is fitted well by a line $\eta / B_0^2 = \text{constant}$, which was also found by \citet{tobias:2007}.  The parameters of Figures~\ref{fig:eta_B0}(a) and (b) are chosen to match those of the simulations performed by \citet{tobias:2007}, the results of which are summarized in Figure~\ref{fig:etaB0_Tobias} (figure reproduced from paper by \citet{tobias:2007}).  However, we could not match the amplitude and spectral distribution of the input forcing exactly, as these values were not reported in detail.  Despite an imperfect matching of forcings, there is nevertheless remarkable agreement between our findings, which result from examining only the ZI within a quasilinear theory, and the results from the fully nonlinear direct numerical simulations by \citet{tobias:2007}.  Part of the reason for this success is that within the ZI calculation, the details of the forcing turn out not that important. As we have argued in Sections~\ref{sec:Orr} and~\ref{sec:ZI}, zonal jet appearance is controlled by the competition between the drive (Reynolds stresses) and suppressor (Maxwell stresses). The amplitude of the forcing, though, does not control this difference since both Reynolds and Maxwell stresses are proportional to the total energy input rate by the forcing.  For example, compare Figures~\ref{fig:eta_B0}(a) and~\ref{fig:eta_B0}(b), which use the same input parameters except for a forcing strength that differs by two orders of magnitude.  Qualitatively and quantitatively, the zonation boundary separating robust zonal flow growth (plus signs) from other behavior (circles and asterisks) changes little. 



Also shown in each plot of Figure~\ref{fig:eta_B0} is a black contour, which depicts the curve $(\wA^2 / \wR^2) (1 + \Prm)^2 / \Prm = 1$.  To compute a single number for $\wA^2 / \wR^2$, we use a characteristic wavenumber, which we take to be the forcing wavenumber $k_f$.  In the regime $\Prm \gg 1$, or $\n \gg \eta$ (the bottom half of the curve), this curve reduces to $(\wA^2 / \wR^2) \Prm = 1$.  This equation recovers the observed scaling $B_0^2  / \eta = \text{constant}$, but also provides a value for the constant.  As seen in Figure~\ref{fig:eta_B0}, this constant works remarkably well at disparate values of $\n$ (separated by four orders of magnitude) at determining the $\eta / B_0^2$ boundary.

The parameter
\begin{equation}
	\Y \equiv (\wA^2 / \wR^2) (1 + \Prm)^2 \big/ \Prm,
	\label{eq:upsilon}
\end{equation}
is derived from the level of magnetic fluctuations in the homogeneous equilibrium $G^H$. The expression is given in \eref{eqs:cov_hom}.  A key parameter determining the homogeneous equilibrium is
	\begin{equation}
		z \defineas \wR^2 + (\n + \eta)^2 k^4 + \frac{(\n + \eta)^2}{\n \eta} \w_A^2.
	\end{equation}
In the regime of $\n k^2,\; \eta k^2,\; \wA \ll \wR$, the middle term of $z$ is negligible.  The third term can be large or small compared to $\wR^2$ because $(\n + \eta)^2 / \n \eta = (1 + \Prm)^2 / \Prm $ can be big if either of $\n$ or $\eta$ is much larger than the other.  The critical parameter $\Y$ is the ratio of the third term to the first term.  If $\Prm$ is not too large or too small such that $\Y \ll 1$, then $z \approx \wR^2$.  Furthermore, if the additional assumption is made that $\Prm \gg 1$ but still $\Y \ll 1$, the covariance of magnetic fluctuations becomes, from \eref{eqs:cov_hom_G},
	\begin{equation}
		\hat{G}^H_{\v{k}} = \frac{\wA^2}{\wR^2} \frac{\hat{Q}_{\v{k}}}{2 \eta k^6}.
	\end{equation}
Hence, the covariance of magnetic fluctuations scales as $B_0^2 / \eta$, while in the same regime, the covariance of hydrodynamic fluctuations $\hat{W}^H$ is independent of both $\eta$ and $B_0$.  Thus, we have related parameters that determine the magnetic fluctuation level to the boundary of zonostrophic instability and found good agreement.  The precise physics determining $\Y = 1$ as a critical value  (when $\Prm \gg 1$) are not fully understood.  However, the agreement between $\Y=1$ and the zonostrophic instability boundary is broadly consistent with the idea that magnetic fluctuations oppose zonostrophic instability, and hence suppress zonal flow.

\section{Discussion}
We have presented a theoretical explanation for the zonal flow suppression previously observed in simulations that imposed a background magnetic field aligned with the direction of rotation.  Our calculations show that the Maxwell stress, caused by magnetic fluctuations, tends to suppress the instability that leads to zonation.  We have performed two separate calculations: a simple calculation based on the Kelvin--Orr shearing wave and a more elaborate calculation based on the CE2 statistical framework. We found consistent results.

We summarize our findings as follows.
\begin{enumerate}
	\item  We have generalized the Kelvin--Orr shearing wave dynamics to include magnetic fields.  In a decomposition into the natural modes of the system, the fast and slow magneto-Rossby waves, we found that the fast wave, which reduces to the Rossby wave for a vanishing magnetic field, can drive and reinforce a weak mean zonal flow.  The slow wave opposes the growth of a weak flow.
	\item We have generalized the zonostrophic instability to include magnetic fields.  In the limit of long-wavelength weak mean flow with weak background magnetic field, the physics of the Kelvin--Orr shearing wave dynamics is recovered.
	\item We demonstrated that the background magnetic field suppresses formation of zonal flow by quenching the instability of initial growth rather than through other means.  (For example, it could have been the case that magnetic fields destabilized finite-amplitude mean flows.)
	\item We showed that a background magnetic field can suppress the formation of zonal flows even when $\w_A^2 \ll \w_R^2$.  This occurs because strong magnetic fluctuations can develop. These magnetic fluctuations give rise to a Maxwell stress that opposes the Reynolds stress that was reinforcing weak shear flows.  This is consistent with the numerical results of \citet{tobias:2007}.
	\item In the regime $\n k^2, \eta k^2, \w_A \ll \w_R$, the quasilinear prediction of zonostrophic instability and the results of fully nonlinear direct numerical simulations by \citet{tobias:2007} are in good agreement for predicting the boundary in parameter space where zonation occurs.
\end{enumerate}

We found that suppression of zonostrophic instability occurs for two reasons.  First, the stronger the magnetic field, the greater fraction of the total fluctuation energy partitions into magnetic energy as opposed to hydrodynamic energy.  Hence, turning up the magnetic field decreases the relative strength of the Reynolds stress, which drives zonal flow, and increases the strength of the Maxwell stress, which suppresses zonation.  Second, increasing the magnetic field modifies the eigenmode character of the fast and slow waves.  The fast wave changes from a Rossby wave at $B_0 = 0$ to an {\Alfven} wave at large $B_0$.  We found that the fast wave's contribution to driving a mean flow decreases as $B_0$ increases.

In this regime, we have mostly focused on ($\n k^2$, $\eta k^2$, $\w_A \ll \w_R$); the former mechanism is the effective one because it leads to zonation suppression for even relatively weak magnetic fields.  For instance, Figure~\ref{fig:eta_B0} shows that magnetic suppression of zonal flow can occur even for $\w_A / \w_R \lesssim 10^{-2}$ as long as $\eta$ is sufficiently small.  In contrast, for the latter mechanism to have an appreciable effect, the magnetic field must be sufficiently strong such that {\Alfven} frequency is comparable to or larger than the Rossby frequency.

We note that although it has been suggested to examine the {\Alfven} wave properties calculated from the total magnetic field (background \& perturbed; \citet{tobias:2007}), within the quasilinear dynamics used in this study, only the background magnetic field $B_0$ determines the {\Alfven} wave properties.

We now turn to discussion of two assumptions used in both the Kelvin--Orr and the ZI calculations that at first glance appear incompatible.  First, we have neglected eddy--eddy nonlinearities.  Second, we have assumed a very weak shear flow.  It is true that both of these assumptions cannot be quantitatively satisfied.  However, the question that primarily concerns us here is can we understand some physics with these assumptions?  We think the answer is yes.  The calculations under these assumptions reveal a coherent effect in which fluctuations are organized by a shear flow to either reinforce or oppose that shear flow.  Qualitatively, one could see how this same coherent effect could occur even without neglecting eddy--eddy nonlinearities, which may be more incoherent in nature and not disrupt the coherent process. 

The eddy-mean flow interaction between the coherent flow and the incoherent eddy field is so robust that it manifests itself even when the mean flow is weak. This fact has been revealed in previous studies of unmagnetized flows \citep{bakas:2013-prl,constantinou:2014}. For example, \citet{constantinou:2014} compared predictions of ZI with fully nonlinear direct numerical simulations and showed that the bifurcation to zonation (i.e.,~when zonal flows are still very weak) is indeed well captured in the quasilinear model, so long as the eddy field is modified to match that in nonlinear simulations. Here, the agreement of the magnetized ZI with the simulations results by \citet{tobias:2007} indicates that in magnetized fluids, the eddy-mean flow interaction retained within the quasilinear approximation is the dominant process responsible for driving or opposing zonal flows.

In conclusion, we have explained how magnetic fields can suppress zonation in a rotating MHD fluid through a relatively simple mechanism.  In the absence of a magnetic field, an initially weak shear flow organizes hydrodynamic fluctuations to reinforce itself and grow.  But in the magnetized case, a weak shear flow coherently organizes magnetic fluctuations to oppose it.

\acknowledgments
We would like to thank the organizers of the workshop ``Vorticity in the Universe,'' which was held in the Aspen Center for Physics, 2017 August 27th--September 17th. This work was performed, in part, at the Aspen Center for Physics, which is supported by the National Science Foundation grant PHY-1607611. We also thank Petros Ioannou and Steve Tobias for fruitful discussions.  This work was performed under the auspices of the U.S.\ Department of Energy by Lawrence Livermore National Laboratory under contract No.~DE-AC52-07NA27344.  N.C.C.~also acknowledges partial support from the National Science Foundation under Award OCE-1357047.

\appendix

\section{Zonostrophic instability with magnetic field}
\label{app:zonostrophic}

The CE2 system of Eqs.~\eqref{eq:ubar}, \eqref{eq:Abar}, and \eqref{eqs:covar} possesses an equilibrium that is statistically homogeneous in both dimensions.  The equilibrium consists of zero mean fields ($\bar{u}=0$, $\bar{A}=0$) and eddy covariances that are determined by a balance of forcing and dissipation.  We perturb about this equilibrium to derive the dispersion relation for growth of mean fields in the zonostrophic instability.

The homogeneous equilibrium covariances can be expressed in terms of their Fourier transforms,~e.g.,
\begin{equation}
  W^H = \sum_{\v{k}} \hat{W}_{\v{k}}^H e^{i\v{k}\bcdot(\v{x}_a-\v{x}_b)} ,
\end{equation}
and similarly for $M^H$, $N^H$, and $G^H$.  From Eqs.~\eqref{eqs:covar} and \eqref{eqs:Lops}, the homogeneous equilibrium can be found to be
	\begin{subequations}
	\label{eqs:cov_hom}
	\begin{align}
  	\hat{W}^H_{\v{k}} &= \frac{ \eta  k^8 (\eta +\nu )^2+ \eta k_x^2\beta ^2  + (\eta +\nu ) B_0^2 k_x^2 k^4  }{ \nu \left[ \eta  k^8 (\eta +\nu )^2 + \eta \beta ^2   k_x^2\right]+ (\eta +\nu )^2 B_0^2 k_x^2 k^4 } \frac{\hat{Q}_{\v{k}}}{2k^2}  ,\\
 		\hat{M}^H_{\v{k}} &= \frac{ -i \eta B_0  k_x \left[ k^4 (\eta +\nu )+ i \beta  k_x\right]}{\nu \left[ \eta  k^8 (\eta +\nu )^2 + \eta \beta ^2   k_x^2\right]+ (\eta +\nu )^2 B_0^2 k_x^2 k^4 }\frac{\hat{Q}_{\v{k}}}{2k^2} ,\\
 		\hat{N}^H_{\v{k}} &= (\hat{M}^H_{\v{k}})^* ,\\
 		\hat{G}^H_{\v{k}} &= \frac{(\eta +\nu )B_0^2 k_x^2}{\nu \left[ \eta  k^8 (\eta +\nu )^2 + \eta \beta ^2   k_x^2\right]+ (\eta +\nu )^2 B_0^2 k_x^2 k^4 }\frac{\hat{Q}_{\v{k}}}{2k^2} , \label{eqs:cov_hom_G}
	\end{align}
	\end{subequations}
with $k\equiv|\v{k}|$. Note that, in general, property~\eref{eq:symm_MN} together with the fact that both $M$ and $N$ are real implies that $\hat{N}^H_{\v{k}} = (\hat{M}^H_{\v{k}})^*$. The stresses in~\eref{eq:stresses} that correspond to~\eref{eqs:cov_hom} are exactly zero, a consequence of statistical homogeneity in the $y$ direction.

We perturb the homogeneous equilibrium as $\bar{u}=\delta\bar{u}$, $\bar{A}=\delta\bar{A}$, $W=W^H + \delta W$, etc, and substitute into the linearized CE2 equations. The perturbations are Fourier-decomposed as
	\begin{subequations}
	\label{eqs:s3teigen}
	\begin{align}
		\delta \bar{u}  &= c_u\,e^{\l t} e^{i q y}, \\
		\delta \bar{A}  &= c_A\,e^{\l t} e^{i q y} ,\label{eq:dudA_ansatz} \\
		\delta W &= e^{\l t} e^{i q(y_a+y_b)/2} \sum_{\v{k}}  \hat{w}_{\v{k}}\,e^{i\v{k}\bcdot(\v{x}_a-\v{x}_b)} ,
	\end{align}
	\end{subequations}
and similarly for $\delta M$, $\delta N$, and $\delta G$.  Here, $\l$ is the eigenvalue and $q$ is the perturbation wavenumber of the zonal flow.

We describe the procedure for the rest of this calculation as follows. We insert \eref{eqs:s3teigen} into the linearized CE2 equations and solve for $\hat{w}_{\v{k}}$, $\hat{m}_{\v{k}}$, $\hat{n}_{\v{k}}$, and $\hat{g}_{\v{k}}$ as functions of $c_u$, $c_A$, $\l$, $q$ and the equilibrium covariance spectra (Eqs.~\eqref{eq:ch_matrix1}, \eqref{eq:ch_matrix2}).  Having $\hat{w}_{\v{k}}$, $\hat{m}_{\v{k}}$, $\hat{n}_{\v{k}}$, and $\hat{g}_{\v{k}}$ in hand, we derive expressions for the stresses (which again depend on $c_u$, $c_A$, $\l$, and $q$; see Eqs.~\eqref{eqs:eigenstresses}, \eqref{eq:stresses_decomposed}). Then, from the two mean-field perturbation equations we end up with a linear system for $c_u$ and $c_A$ (\eref{eq:linearsystem}) that has non-trivial solutions only for particular values of $\l$ (\eref{eq:eigenvalue}).

After substitution of \eref{eqs:s3teigen}, the perturbation covariance equations can be placed into the form
\begin{align}
\mathbb{F} \begin{pmatrix} \hat{w}_{\v{k}}\\ \hat{m}_{\v{k}}\\ \hat{n}_{\v{k}}\\ \hat{g}_{\v{k}}\end{pmatrix} &= c_u \begin{pmatrix}
			i k_x(1-q^2/k_{1}^2)\hat{W}^H_{\v{k}_{1}}\\
			i k_x\hat{M}^H_{\v{k}_{1}}  \\
			i k_x(1-q^2/k_{1}^2)\hat{N}^H_{\v{k}_{1}}\\
			i k_x\hat{G}^H_{\v{k}_{1}}
	\end{pmatrix} - c_u \begin{pmatrix}
			i k_x(1-q^2/k_{-1}^2)\hat{W}^H_{\v{k}_{-1}}\\
			i k_x(1-q^2/k_{-1}^2)\hat{M}^H_{\v{k}_{-1}}\\
			i k_x\hat{N}^H_{\v{k}_{-1}}\\
			i k_x\hat{G}^H_{\v{k}_{-1}}
		\end{pmatrix} + c_A \begin{pmatrix}
			q k_x(k_{1}^2-q^2)\hat{M}^H_{\v{k}_{1}}\\
			(q k_x/k_{1}^{-2}) \hat{W}^H_{\v{k}_{1}}\\
			q k_x(k_{1}^2-q^2)\hat{G}^H_{\v{k}_{1}}\\
			(q k_x/k_{1}^2) \hat{N}^H_{\v{k}_{1}}
	\end{pmatrix} - c_A  \begin{pmatrix}
		q k_x(k_{-1}^2-q^2) \hat{N}^H_{\v{k}_{-1}}\\
		q k_x(k_{-1}^2-q^2) \hat{G}^H_{\v{k}_{-1}}\\
		(q k_x/k_{-1}^2) \hat{W}^H_{\v{k}_{-1}}\\
		(q k_x/k_{-1}^2) \hat{M}^H_{\v{k}_{-1}}
\end{pmatrix} ,
\label{eq:ch_matrix1}
\end{align}
where
	\begin{align}
		\mathbb{F} \equiv \begin{pmatrix}
			\l + \nu \left( k_{1}^2+ k_{-1}^2\right) + 2i \b k_x k_y q\big/(k_{1}^2 k_{-1}^2)   &  +i k_x B_0k_{-1}^2 & -i k_x B_0k_{1}^2 & 0\\
			 +i k_x B_0\big/k^2_{-1}  &  \l +\nu k_{1}^2+\eta k_{-1}^2 - i \b k_x\big/k_{1}^2   & 0 & -i k_x B_0 k_1^2\\
			-i k_x B_0\big/k^2_{1}  &  0 & \l+ \eta k_{1}^2 + \nu k_{-1}^2 + i \b k_x\big/k_{-1}^2 &  +i k_x B_0 k_{-1}^2\\
			0  &  -i k_x B_0\big/k^2_{1} & +i k_x B_0\big/k^2_{-1}&\l+\eta( k_{1}^2+ k_{-1}^2)
	\end{pmatrix}.
	\label{eq:ch_matrix2}
	\end{align}
Above, we used the notation $\v{k}_{\pm 1} \equiv (k_x,k_y\pm q/2)$, and $k_{\pm 1} = |\v{k}_{\pm 1}|$. Note that it is important to keep both $\delta M$ and~$\delta N$; we cannot use the property~\eref{eq:symm_MN} to relate $\hat{n}_{\v{k}}$ to $\hat{m}_{\v{k}}$ here because the perturbations $\delta M$ and $\delta N$ have been represented with a complex eigenfunction.  Equation \eqref{eq:ch_matrix1} relates the eigenmode components $\hat{w}_{\v{k}}, \hat{m}_{\v{k}}, \hat{n}_{\v{k}}, \hat{g}_{\v{k}}$ and $c_u, c_A$ in a matrix equation.

What we would like is to write each of $\hat{w}_{\v{k}}$, etc., in terms of $c_u$ and $c_A$.  To do so, we invert the system \eqref{eq:ch_matrix1}, or equivalently, invert $\mathbb{F}$, using the computer algebra system Mathematica.  The resulting expressions for $\hat{w}_{\v{k}}$, etc., are extremely complicated and so they are not written explicitly.  We note that $\hat{w}_{\v{k}}$, etc., are linear in both $c_u$ and $c_A$.

With $\hat{w}_{\v{k}}$, $\hat{m}_{\v{k}}$, $\hat{n}_{\v{k}}$, and $\hat{g}_{\v{k}}$, we can write the perturbation stresses as
	\begin{subequations}
	\label{eqs:eigenstresses}
	\begin{align}
		\delta\overline{v'\z'} &= e^{i q y} \sum_{\v{k}} \frac{i q k_x k_y}{k_1^2 k_{-1}^2} \hat{w}_{\v{k}}, \label{eq:dvz}\\
		\delta\overline{(\partial_x A')\nabla^{2} A'} &= e^{i q y}  \sum_{\v{k}} i q k_x k_y\, \hat{g}_{\v{k}}, \label{eq:dAlapA}\\
		\delta\overline{v'A'} &= e^{i q y} \sum_{\v{k}} i k_x \left(-\frac{\hat{m}_{\v{k}}}{2k_{1}^2} +\frac{\hat{n}_{\v{k}}}{2k_{-1}^2}\right).\label{eq:dvA}
	\end{align}
	\end{subequations}
To obtain the above we used Eqs.~\eqref{eq:stresses} and \eqref{eqs:s3teigen}.  Since $\hat{w}_{\v{k}}$, etc. are linear in $c_u$ and $c_A$, it is useful to decompose the stresses as
	\begin{subequations}
	\label{eq:stresses_decomposed}
	\begin{align}
		\delta\overline{v'\z'} &=  e^{i q y} \left[ c_u \delta\overline{v'\z'}^u + c_A \delta\overline{v'\z'}^A\right]  ,\label{eq72}\\
		\delta\overline{(\partial_x A')\nabla^{2} A'} &= e^{i q y} \left[ c_u\,\delta\overline{(\partial_x A')\nabla^{2} A'}^u + c_A\,\delta\overline{(\partial_x A')\nabla^{2} A'}^A\right] ,\label{eq73}\\
		\delta\overline{v'A'} &= e^{i q y} \left[ c_u \delta\overline{v'A'}^u + c_A \delta\overline{v'A'}^A\right] .\label{eq74}
	\end{align}
	\end{subequations}
Explicit expressions for the terms such as $\delta\overline{v'\z'}^u$ are derived, but again are too complicated and unilluminating to include here.  Substituting~\eref{eq:stresses_decomposed} into the linearized mean-field equations, we obtain the linear system of just two equations
	\begin{subequations}
	\label{eq:linearsystem}
	\begin{gather}
		c_u \left[ \l + \nu q^2 - \delta\overline{v'\z'}^u + \delta\overline{(\partial_x A')\nabla^{2} A'}^u \right] + c_A \left[ \delta\overline{(\partial_x A')\nabla^{2} A'}^A-\delta\overline{v'\z'}^A \right]=0  ,\label{eq:dpsim}\\
		c_u \left[ i q \, \overline{v'A'}^u \right]  + c_A \left[ \l + \eta q^2 + i q\, \delta\overline{v'A'}^A \right] = 0 .\label{eq:dAm}
	\end{gather}
	\end{subequations}
Equation \eqref{eq:linearsystem} has a non-trivial solution only if
	\begin{equation}
		\left[\l + \nu q^2 - \delta\overline{v'\z'}^u + \delta\overline{(\partial_x A')\nabla^{2} A'}^u\right]\left[\l + \eta q^2 + i q \, \overline{v'A'}^A \right] - i q \, \overline{v'A'}^u \left[ \delta\overline{(\partial_x A')\nabla^{2} A'}^A - \delta\overline{v'\z'}^A\right] = 0 .\label{eq:eigenvalue}
	\end{equation}
Equation~\eqref{eq:eigenvalue} is a single nonlinear equation that determines the allowed eigenvalues $\lambda$.  We solve it with Newton's method.  We typically must scan over various initial guesses to ensure we do not miss an unstable eigenvalue.  With the eigenvalue in hand, we can return to~\eref{eq:linearsystem} and compute the coefficients $c_u$ and $c_A$.

A more straightforward way to perform the ZI analysis is to write explicitly the matrix that governs the linearized dynamics of the full state vector (i.e., for $\delta\bar{u}$, $\delta\bar{A}$ and for \emph{all} wavenumber components of $\hat{w}_{\v{k}}$, $\hat{m}_{\v{k}}$, $\hat{n}_{\v{k}}$, and~$\hat{g}_{\v{k}}$), and then perform eigenanalysis of this matrix numerically.  The resulting matrix can be somewhat large, but it is still feasible to directly compute all eigenvalues.  In this method, one does not have to worry about missing any eigenvalues or about the initial guess to provide to the Newton solver.

In this paper we have performed the stability calculations using both methods and found exactly the same results.  The former method, which uses the inversion of~$\mathbb{F}$, is particularly useful for analytical insight. For example, Figure~\ref{fig:stresses_qphigrid} relies on the inversion of~$\mathbb{F}$.  Additionally, the former method enables an asymptotic expansion of the expression for the stresses that recovers the same parameter dependence found in the Kelvin--Orr shearing wave calculation, as discussed in Section~\ref{sec:ZIexample}.

\end{document}